\def\half{\frac{1}{2}}

\def\beqar{\begin{eqnarray}}
\def\eeqar{\end{eqnarray}}

\def\dt{\hat\partial_o}
\newcommand{\llabel}[1]{\label{#1}}              % DO NOT show equation label

\newcommand{\labeq}[2]{ \begin{equation} \llabel{#1}
{#2}
\end{equation}}

\documentclass[aps,showpacs,twocolumn,prd,preprintnumbers,amsmath,amssymb,letterpaper]{revtex4}

\usepackage{amsmath}

\begin{document}

\title{\bf Mixed Hyperbolic - Second-Order Parabolic Formulations of General Relativity}

\author{Vasileios Paschalidis${}^1$}

\affiliation{${}^1$ Department of Astronomy and Astrophysics, The
University of Chicago, 5640 S Ellis Ave., Chicago IL 60637}

\date{\today}

\begin{abstract}

Two new formulations of general relativity are introduced. The first one is a parabolization of the Arnowitt, Deser, Misner (ADM) formulation 
and is derived by addition of combinations of the constraints and their derivatives to the right-hand-side of the ADM
evolution equations. The desirable property of this modification is that it turns the surface of constraints into a local attractor because the constraint propagation equations become second-order parabolic independently of the gauge conditions employed. This system may be classified as mixed hyperbolic - second-order parabolic. The second formulation is a parabolization of the Kidder, Scheel, Teukolsky  formulation and is a manifestly mixed strongly hyperbolic - second-order parabolic set of equations, bearing thus resemblance to the compressible Navier-Stokes equations. As a first test, a stability analysis of flat space is carried out and it is shown that the first modification exponentially damps and smoothes all constraint violating modes. These systems provide a new basis for constructing schemes for long-term and stable numerical integration of the Einstein field equations.

\end{abstract}

\pacs{04.25.Dm, 04.70.Bw}

%\keywords{}

\maketitle


%\numberwithin{equation}{section}

%%%%%%%%%%%%%%%%%%%%%%%%%%%%%%%%%%%%%%%%%%%%%%%%%%%%%%%%%%%%%%%%%%%%%
\section{Introduction}
%%%%%%%%%%%%%%%%%%%%%%%%%%%%%%%%%%%%%%%%%%%%%%%%%%%%%%%%%%%%%%%%%%%%%

For years people have tried to obtain analytic solutions of the complex field equations of Einstein's general theory of relativity (GR). Apart from few cases where symmetry is invoked, it is almost impossible to analyze the complicated dynamics in the strong gravitational field regime as that is described by GR. Approximation methods have been developed over the course of time, but the most promising tool for tackling problems such as gravitational waves arising from binary black hole (BBH) or binary neutron star (BNS) mergers, gravitational collapse etc., is numerical relativity. 

In the past few years remarkable progress has been made towards achieving
long term and stable evolution of the Einstein field equations.  Recently 
particular cases of the long standing problem of the evolution 
of BBH were solved \cite{Pret, Camp, Baker}. Despite this exceptional achievement,
the general problem of long term and stable evolution of the GR equations remains open and there is still much work left to be done. There is 
 no theory or prescription to chose what formulation(s) and 
what gauge conditions are suitable for the numerical solution of a given problem. For example, there does not seem to exist a definitive explanation of why 
the approaches of \cite{Pret, Camp, Baker} perform so well when contrasted to previous efforts. Furthermore, there are astrophysical and theoretical problems of great interest for which the formulations above have not been applied yet and it is not known whether they will prove successful in such cases. Such problems are  the study of the internal structure of black holes and astrophysical phenomena, where except for black holes matter is also involved.

The numerical integration of the Einstein equations is not an easy task, because the computations can become unstable and an exponential blow up of the numerical error may occur, even when the formulation employed admits a well-posed initial value problem.  If the numerical techniques, gauge and boundary conditions chosen do not suffer by pathologies, perhaps the most important cause of instabilities during a free evolution is the growth of the constraint violating modes which are excited by numerical errors.

Several methods have been proposed to deal with this last type of instability. A path followed by many relativists is referred to as constrained evolution \cite{Stark, Abrahams92, Choptuik93, Abrahams93, Abrahams94, Choptuik03a, Choptuik03b}, in which a subset of the dynamical variables is evolved in time by use of the evolution equations and the remaining variables are obtained by solving the constraints after each time-step. This method has proven to be very accurate and has been successfully applied in cases where some type of symmetry is present. Another method, which resembles the approach of constrained evolution, is called constraint projection \cite{Holst,Anderson}. In this method all variables are evolved in time using the evolution equations and they are periodically ``projected" via an optimal algorithm to the closest configuration of dynamical fields which satisfies the  constraints. This method is quite robust, but it can be computationally demanding because it requires the repeated solution of non-linear elliptic equations.

Another strategy takes advantage of the fact that in the ideal case where the constraint equations are satisfied, one has the freedom to add combinations of the constraints to the right-hand-side (RHS) of the evolution equations. This approach changes the mathematical properties of the equations, but not the physics these equations describe. A number of strongly and symmetric hyperbolic evolution equations have been constructed using this method, see \cite{YonedaShinkai} for a review, which resulted in improved stability of numerical simulations of GR. 

By virtue of the freedom to add the constraint equations to the RHS of the evolution equations and/or by extension of the number of dynamical fields, it is also possible to add terms that act as ``constraint drivers", i.e., terms which turn the constraint surface into a local attractor. This technique is nowadays usually termed as ``constraint damping". 

The constraint damping strategy naturally splits in two different categories: a) Additional, non-physical variables are appropriately introduced in the evolution system, so that their time evolution drives the physical variables toward the surface of constraints \cite{Brodbeck,Gundlach}, and b) no additional (non-physical) variables are introduced in the system, but only combinations of the constraints, which act as constraint drivers, are added to the RHS of the evolution equations of the physical fields, e.g. \cite{Robert_Owen}. In this case the constraint dissipation is inherent to the evolution equations of the physical variables themselves.

The goal of many formulations in numerical relativity has been to incorporate such drivers in symmetric or strongly hyperbolic systems without changing the principal part of the evolution equations \cite{Brodbeck, Gundlach}. The common property of all these formulations is that the damping rate of the short-wavelength constraint-violating modes is independent of the wavelength of these modes. As a result, the constraint violations are not smoothed during a free evolution, unless numerical dissipation is artificially added to a code. The existence of short-wavelength components may lead to spurious oscillations and potentially  terminate the computations \cite{Pret}. 

For this reason, it may be advantageous to consider formulations of GR which possess wavelength-dependent constraint damping. One way to achieve this goal is to construct formulations of GR that under free evolution force all or some of the constraints to evolve according to parabolic equations. Parabolic equations damp short wavelengths extremely efficiently and for this reason they posses desirable smoothing properties  \cite{PDE_textbook}.

The purpose of this paper is to present two new parameterized  3+1 formulations which fall in category b) of the constraint damping strategy. These two formulations will be referred throughout this work as system 1 and system 2. The difference between these two systems and most formulations of GR, which belong to the same category, is that constraint dissipation does not only enter through low order terms, but also through higher order terms since the order of the of the evolution equations is increased. In system 1 and system 2 all or some of the constraint violating modes have been parabolized.

The desirable property of the well-posed system 1 is that the evolution equations of the constraints become second-order parabolic, independently of the gauge conditions and spacetime configuration, which in turn implies that the constraint surface becomes a local attractor. The evolution equations of system 1 may be classified as a set of mixed hyperbolic- second-order parabolic partial differential equations (PDEs).

The well-posed system 2, is a parabolization of what is usually referred to as 
the Kidder, Scheel, Teukolsky (KST) formulation which was introduced in \cite{KST} and later extended in \cite{KST2}. A generalization of the gauge condition chosen in \cite{KST} is given and a certain combination of the terms of system 1 is added to the RHS of the KST formulation. Unlike system 1, system 2 forces only a subset of the constraint variables to evolve according to parabolic equation and the constraint propagation system is mixed strongly hyperbolic - second-order parabolic (MHSP).

It is here noted that hyperbolic and parabolic drivers have been constructed for the MHD equations, by introducing additional non-physical fields, whose time evolution drive the physical fields on the constraint surface, see for example \cite{Dedner} and references therein. A similar approach has also been applied in numerical relativity via a Hamiltonian constraint relaxation method \cite{Marronetti}, which introduces a parabolic equation to approximately solve the Hamiltonian constraint. The parabolic equation is solved not in real (coordinate) time, but in some fiducial time until the Hamiltonian constraint has been relaxed. The difference between the suggested approach in this work and the aforementioned two efforts is that here there is no need for extra variables and that the constraint damping is inherent to the evolution equations of the physical dynamical variables, which implies that the damping takes place in real time and not in some fiducial time. In addition to that, in contrast to the approach of \cite{Marronetti}, system 1 dissipates the perturbations which violate not only the Hamiltonian constraint, but all constraints of the formulation.

Note also that a general, robust constraint driver has been developed and successfully applied to two formulations of Maxwell's equations in \cite{FiskePRD}, where the evolution equations were cast into a type of MHSP set of equations. In \cite{FiskeThesis}, the same approach has been successfully applied to a ``1+1" case of the ADM evolution equations, linearized about flat space, where the resulting evolution system resembled the structure of a mixed hyperbolic - fourth-order parabolic set of PDEs. The approach of \cite{FiskePRD} can in general be applied to the full non-linear equations of the first-order ADM formulation, which is re-introduced in this paper (overcoming this way the obstacle of fourth-order evolution systems). However, the main disadvantage of this method is that the resulting evolution system would turn from quasi-linear into highly non-linear, see \cite{FiskeThesis}, because the constraints would be added in quadrature. On the other hand, the ideas suggested in this work result in quasi-linear PDEs and this constitutes one of the major differences between this paper and the work in \cite{FiskePRD}. 

Finally, it must be mentioned that the idea of adding multiples of the derivatives of the constraints to the RHS of the evolution equations has been independently suggested in \cite{Calabrese}, where the approach was applied successfully to different formulations of Maxwell's equations.

This paper is organized as follows. In section~\ref{model_equation}, a model equation is presented to illustrate the parabolization approach. 
The first order ADM formulation and gauges are presented in section~\ref{ADM_formulation}. In section \ref{system_1} system 1 is presented and studied. A review of the KST formulation is presented, a generalization of the gauge of the original KST formulation is made and system 2 is given and studied in section \ref{system_2}.  In section \ref{flat_space_stability} a study of the stability properties of flat space with three different formulations is presented. In section \ref{parameters} a brief discussion of the choice of parameters is given. Finally, the work is concluded in section~\ref{discussion}.

%%%%%%%%%%%%%%%%%%%%%%%%%%%%%%%%%%%%%%%%%%%%%%%%%%%%%%%%%%%%%%%%%%%%%%%%%%%%%%%%%%%%%%%%%%%%
\section{A simple model equation \label{model_equation}}
%%%%%%%%%%%%%%%%%%%%%%%%%%%%%%%%%%%%%%%%%%%%%%%%%%%%%%%%%%%%%%%%%%%%%%%%%%%%%%%%%%%%%%%%%%%%

To illustrate the strategy that will be employed in order to  modify the Einstein equations a scalar wave equation for the variable $g(x,t)$ is introduced. A linear one-dimensional scalar wave equation is given by
\labeq{wave_equation}{
\partial_t\partial_t g=\partial_x\partial_x g,
}
where $\partial_x$ denotes a spatial derivative and $\partial_t$ a time derivative. 
Equation \eqref{wave_equation} is equivalent to the following first-order system 
\labeq{wave_equation_1st}{\begin{split}
\partial_t g=& K, \\
\partial_t K=&  \partial_x D, \\
\partial_t D=& \partial_x K.
\end{split}
}
Equations \eqref{wave_equation_1st}  have the same totality of solutions as equation \eqref{wave_equation} only if the following constraint is 
satisfied at all times
\labeq{constraint_wave_equation}{
{\cal C}\equiv\partial_x g-D=0.}
Equation \eqref{constraint_wave_equation} is essentially the definition of the $D$ variable.
In addition to the evolution equations of the dynamical variables $g,K,D$, one can also obtain the evolution equation of the constraint ${\cal C}$. By taking a time derivative of the constraint one finds
\labeq{constraint_wave_evol}{
\partial_t {\cal C}=0.}

The initial value problem in GR is similar to that of equations \eqref{wave_equation_1st} and \eqref{constraint_wave_equation}, in the sense that there is a set of evolution equations for the dynamical variables and a set of constraints that have to be satisfied at all times. In GR however, the equations are so much more complex that during their numerical integration the constraints may grow exponentially with time leading the computations to a termination. 

This is not the case with equation \eqref{wave_equation_1st}, because 
\eqref{wave_equation_1st} is a set of  strongly hyperbolic PDEs (definitions and theorems pertaining to hyperbolic PDEs can be found in  \cite{PDE_textbook}) and hence a perfectly well-behaved one. It can be solved numerically, without any difficulty in satisfying the constraint \eqref{constraint_wave_equation} at all times.  Therefore, the purpose of the model equation is to demonstrate a strategy in order to  suppress very efficiently the growth of constraint violating modes. 

This strategy is based on the fact that not only the constraint ${\cal C}$ has to be zero at all times, but also its spatial derivative. This implies that one can add combinations of those two to the RHS of the evolution equations \eqref{wave_equation_1st}, without changing the solutions of the equations. With that in mind, the following modification is suggested in order to damp any constraint violating modes and still retain the well-posedness of the system.
\labeq{modified_wave_equation}{\begin{split}
\partial_t g=& K +\lambda\partial_x {\cal C},\\
\partial_t K=&  \partial_x D, \\
\partial_t D=& \partial_x K+\zeta C.
\end{split}
}
In order to prove that  \eqref{modified_wave_equation} admits a well-posed Cauchy problem, it is instructive to present from \cite{PDE_textbook} some definitions and theorems concerning parabolic and mixed parabolic-hyperbolic systems of quasi-linear partial differential equations .

Consider a quasi-linear even-order operator of the form
\labeq{even_op}{
P(\partial_x)=P_{2m}(\partial_x)+Q(\partial_x),}
where 
\labeq{P2m}{P_{2m}(\partial_x)=\sum_{|\nu|=2m}A_{\nu}(u) \partial_1{}^{\nu_1}\partial_2{}^{\nu_2}\cdots \partial_d{}^{\nu_d},
}
and
\labeq{Qm}{Q(\partial_x)=\sum_{|\nu|\leq 2m-1}A_{\nu}(u) \partial_1{}^{\nu_1}\partial_2{}^{\nu_2}\cdots \partial_d{}^{\nu_d},
}
where $\sum_{i=1,d} \nu_i=\nu$, $\partial_s{}^{\nu_s}$ denotes the partial derivative with respect to $x^s$ of order ${\nu_s}$, 
$s=1,\ldots, d$ with $d$ the  number of spatial dimensions, $u$ is the column vector of the $n$ dynamical variables and $A_{\nu}$ are $n\times n$ matrix functions of the dynamical variables, but not their derivatives.

\paragraph{\textbf{Definition 1.}}
The equation $\partial_t u=P(\partial_x) u$ is called parabolic, if for all $\kappa\epsilon {\bf R}^d$ the eigenvalues 
$\omega_j(\kappa), \ \ j=1,\ldots, n$ of $P_{2m}(i \kappa)=-\sum_{\nu=2m}A_{2m}\kappa_1{}^{\nu_1}\cdots \kappa_d{}^{\nu_d}$ satisfy
\labeq{parabolic_condition}{Re(\omega_j(\kappa))\leq -\delta |\kappa|^{2m}, \ \ j=1,\ldots, n,}
with some $\delta >0$ independent of $\kappa$, where in the symbol $P_{2m}(i\kappa)$, $i^2=-1$.
%
%\paragraph{\textbf{Definition 2.}} 
%A second order systems of the form
%%\labeq{strongly_parabolic}{
%\partial_t u=\partial_i(A^{ij}\partial_j u) +Q(\partial_x)u
%}
%
% is called strongly parabolic if 

\paragraph{\textbf{Theorem 1.}}
\textit{The Cauchy problem for a parabolic system $\partial_t u=P(\partial_x) u$ is well-posed}.

Assume now that a system of partial differential equations can be written in the following form
%
%\begin{widetext}
\labeq{MixedPH}{\partial_t\left(
\begin{array}{c}
u \\ 
v  \\
\end{array}\right)=
\left(
\begin{array}{cc}
P_{2} & 0 \\
0 & P_{1} \\
\end{array}\right)
\left(\begin{array}{c}
u \\ 
v 
\end{array}\right)
+\left(
\begin{array}{cc}
0 & R_{12}\\
R_{21} & 0 \\
\end{array}\right)
\left(\begin{array}{c}
u \\ 
v 
\end{array}\right),
%
%+\left(\begin{array}{c}
%G_2 \\ 
%G_1 
%\end{array}\right)
}
%\end{widetext}
where $u$ and $v$ are dynamical variables, $P_2=P_2(\partial_x)$ is a second-order
differential operator, $P_1=P_1(\partial_x)$ is a first-order operator and the coupling terms
$R_{12}=R_{12}(\partial_x)$ and $R_{21}=R_{21}(\partial_x)$ are general first-order differential
operators. Equations \eqref{MixedPH} may or may not have zeroth-order couplings. Further assume that 
the uncoupled systems
\labeq{MHPuncoupled}{\begin{split}
\partial_t u & =P_2(\partial_x)  u \\
\partial_t  v & =P_1(\partial_x) v 
\end{split}
}
are second-order parabolic and first-order strongly hyperbolic, 
respectively. 
\paragraph{\textbf{Definition 2.}} The coupled system \eqref{MixedPH} is called mixed strongly hyperbolic - second-order parabolic.

The coupling terms $R_{12}$ and $R_{21}$ or any low order terms do not destroy the well-posedness of the system and the following theorem can be proved. 
\paragraph{\textbf{Theorem 2.}}
\textit{The Cauchy problem for mixed second-order parabolic-strongly hyperbolic systems \eqref{MixedPH} is well-posed.}

Equation \eqref{modified_wave_equation} is a system of mixed strongly hyperbolic and second-order parabolic PDEs, provided that $\lambda>0$. This can be most easily seen, if the explicit form of \eqref{modified_wave_equation} is written
\beqar
\partial_t g &=& K +\lambda\partial_x\partial_x g-\lambda \partial_x D ,\label{tghyp}\\
\partial_t K&=&  \partial_x D, \label{tKhyp}\\
\partial_t D&=& \partial_x K+\zeta (\partial_x g-D). \label{tDhyp}
\eeqar
It can be readily realized that equations \eqref{tghyp}-\eqref{tDhyp} have the form of
\eqref{MixedPH} with $u=g$ and $v=\{K,D\}$. The coupling terms are $R_{12}=-\lambda\partial_x D$ and $R_{21}=\zeta\partial_x g$ which are first order. Without these first-order couplings, equation \eqref{tghyp} is second-order parabolic when $\lambda>0$, whereas the system of  \eqref{tKhyp} and \eqref{tDhyp} is strongly  hyperbolic.  Therefore, the total set of equations is MHSP and  admits a well-posed Cauchy problem.

 The most interesting aspect of the modification introduced in \eqref{modified_wave_equation} is the structure of the evolution of the constraint ${\cal C}$. A straightforward calculation yields
\labeq{constraint_wave_evol_mod}{
\partial_t {\cal C}=\lambda\partial_x\partial_x {\cal C}-\zeta{\cal C}.}
Equation \eqref{constraint_wave_evol_mod} is  parabolic, provided that $\lambda >0$. Therefore, not only the evolution equations of the dynamical variables are well-posed, but also the evolution of the constraint. 

By Fourier transforming \eqref{constraint_wave_evol_mod}, one can show that damping of the constraint ${\cal C}$ takes place, if
$\lambda>0$ and $\zeta>0$. This is because the damping rate of ${\cal C}$ is of the form $\exp(-(\lambda\kappa^2+\zeta)t)$, where $\kappa$ is the magnitude of the wavenumber. 

By the form of the damping rate of ${\cal C}$ it is evident that the value of $\zeta$ can have a major impact on the long-term stability of the numerical integration of \eqref{modified_wave_equation}, even though theorem 2 dictates that equation \eqref{modified_wave_equation} admits a well-posed Cauchy problem irrespectively of the value of $\zeta$.  

At this point it is instructive to try to understand why this is so. Note that the constraint and its derivative were added to the RHS of equation \eqref{wave_equation_1st}. Note also that the derivative of the constraint $C$ may not be violated, while the constraint itself may be violated, as long as the violation occurs by a constant. Constant violations correspond to infinite wavelength and hence $\kappa=0$. This is the reason why there is no damping, if $\zeta=0$ and $\kappa=0$. To be more specific, at the RHS of the evolution of variable $g$ the derivative of the constraint has been added. So, a violation of the constraint $C$ by a constant is not ``felt" by the $\lambda$ term. The $\lambda$ term damps only the finite wavelength constraint violating perturbations, because $\kappa\neq 0$ perturbations violate not only the constraint, but also the derivatives of the constraint. If $\zeta<0$ then  in order for damping to take place, the following inequality has to be satisfied for all allowed $\kappa$
\labeq{damping_condition}{\lambda\kappa^2>-\zeta.}
In a computer there are limits on the wavenumber of the perturbations, because of the finite length of the computational domain and the mesh size.  The maximum $\kappa$ corresponds  to the Nyquist frequency and the minimum to 
$\kappa_{min}=2\pi/L$, with $L$ the length of the computational domain. Hence, condition \eqref{damping_condition} is satisfied, if $\lambda>-\zeta/\kappa_{min}{}^2$. However, because of the efficiency of the damping very soon all allowed finite wavelength constraint violating perturbations go to zero. As a result the $\lambda$ term in \eqref{modified_wave_equation}  is switched off, since there are no further violations of the derivatives of the constraints. Therefore, the evolution of the constraint $C$ is effectively given by equation 
\eqref{constraint_wave_evol_mod} with $\lambda=0$. If in that case there is any residual $\kappa=0$ violation of the constraints (which is inevitable since an exact zero initial value of a constraint is never achieved in a computer), this violation will obtain explosive behavior (because $\zeta<0$ was assumed), which will occur at a time-scale $\zeta^{-1}$. Therefore, for long-term and stable numerical integration both $\lambda>0$ and  $\zeta>0$ are required. 

A final contradiction has to be resolved here. It was stated earlier that system \eqref{modified_wave_equation} admits a well-posed Cauchy problem irrespectively of the value of $\zeta$, as long as $\lambda>0$. However, in the previous paragraph it was argued that if $\zeta<0$ an exponential blow up of the numerical solution occurs. The resolution of this contradiction takes place, if one realizes that well-posedness does not in general guarantee global in time existence of solutions, but only short time existence. 

The difference between the suggested modification of this section and previous constraint damping strategies is the $\lambda$ term. The advantage of the new approach is that the shortest the wavelength of the constraint-violating perturbations the more efficiently those perturbations are getting damped. And it is the short wavelength perturbations, which produce the most spurious oscillations that may lead to blow up of the numerical error. Of course $\zeta$ is important to damp the long wavelength (including $\kappa=0$) perturbations, which do not get damped very efficiently by the $\lambda$ modification. 

An important result of the wavelength-dependent damping timescale is that the constraint violating noise is smoothed  with time.
Another important property of the suggested modification is that it assigns non-zero ``speed" of propagation to the static mode present in \eqref{wave_equation_1st}. This is may be an advantage because, according to \cite{Alcubierre}, the presence of zero eigenvalues can be responsible for numerical instabilities caused by perturbations by low order terms. 

The same parabolization strategy will be applied to the Einstein equations in section \ref{system_1} and it will be demonstrated that with appropriate modifications all constraints of the theory can be damped.

%%%%%%%%%%%%%%%%%%%%%%%%%%%%%%%%%%%%%%%%%%%%%%%%%%%%%%%%%%%%%%%%%%%%%%%%%%%%%%%%%%%%%%%%%%%%
\section{The first-order ADM formulation \label{ADM_formulation}}
%%%%%%%%%%%%%%%%%%%%%%%%%%%%%%%%%%%%%%%%%%%%%%%%%%%%%%%%%%%%%%%%%%%%%%%%%%%%%%%%%%%%%%%%%%%%

\subsection{3+1 decomposition of spacetime}

The four-dimensional spacetime of the theory
of general relativity can be cast into a 3+1 decomposition \cite{ADM}. This is achieved by assuming that the spacetime can be foliated by a one-parameter 
family of spacelike hypersurfaces. The four-dimensional metric is then written in the following form
\labeq{metric}{ds^2=-\alpha^2dt^2+\gamma_{ij}(dx^i+\beta^i dt)(dx^j+\beta^j dt)
}
where $\gamma_{ij}$ is the positive definite 3-metric on the $t=const.$ hypersurfaces, $\alpha$ is the lapse function, $\beta^i$ the shift vector, and 
$x^i$ are the spatial coordinates, $i=1,2,3$.

To specify the spacetime foliation the gauge equations for the lapse function and the shift vector have to be given. These can generally be written as follows 
\labeq{Gauge}{\begin{split} F_a\bigg (      x^b, \alpha, \beta^i,&
\partial_b\alpha,\partial_b\partial_c\alpha, ..., \partial_b\beta^i
,...
      \gamma_{ij}, \partial_b\gamma_{ij},...
\bigg) =0, \\ & a,b,c = 0,...,3, \quad i,j = 1,2,3. \end{split},} \cite{Khokhlov}.
%\tilde R_{ij}
In this work only algebraic gauges of the following form are considered
\labeq{algebraic_lapse}{\alpha=\alpha(\gamma),}
where $\gamma$ is the determinant of the three-metric $\gamma_{ij}$, and the shift vector $\beta^i$ may be either constant or a fixed
function of the spacetime coordinates ($t,x^i$). In \cite{Paschalidis07, Khokhlov} it was shown that \eqref{algebraic_lapse} makes the evolution
well-posed on the surface of constraints, if 
\labeq{A}{A=\partial\ln\alpha/\partial\ln\gamma>0.}
 Working with \eqref{algebraic_lapse}, one has the flexibility to use either ``1+log" slicing 
\labeq{1pluslog}{\alpha=1+\ln \gamma}
 or a densitized lapse 
\labeq{dense-lapse}{\alpha=Q \gamma^\sigma,}
where $Q$ is constant or a fixed function of $t,x^i$ and $\sigma$ the densitization parameter.
If \eqref{1pluslog} is used, $A=1/\alpha$, whereas if  \eqref{dense-lapse} is employed, $A=\sigma$. Both gauge conditions have 
been used rather successfully in numerical relativity. However, note that ``1+log" slicing 
has better singularity avoidance properties than the densitized lapse \cite{Baumgarte_Shapiro}.

\subsection{Evolution equations and constraints}

The Einstein equations in 3+1 split form are referred to as the ADM formulation and consist 
of two subsets of equations. The first subset is that of the evolution equations, which describe how
the dynamical variables evolve in time. The second subset is that of the constraint equations, which have to be satisfied for
all times. The standard second-order ADM formulation \cite{ADM} has as dynamical variables the 3-metric $\gamma_{ij}$ and the extrinsic curvature $K_{ij}$ of the 3D spacelike hypersurfaces.

The simplest first-order form of the ADM formulation is derived from the second-order one by introducing additional dynamical variables
\labeq{D-defin}{
                      D_{kij} \equiv \partial_k\gamma_{ij},
}
and then deriving the evolution equations for $D_{kij}$. The evolution equations of the first-order ADM are
\labeq{ADM-gamma}{ \hat\partial_o\gamma_{ij} = -2\alpha K_{ij}, }
\labeq{ADM-K}{
\begin{split}
\hat\partial_o K_{ij}  = & -\nabla_i \nabla_j \alpha+\alpha \left(R_{ij} +  K K_{ij} - 2
\gamma^{mn} K_{im} K_{jn} \right),
\end{split}
}
\labeq{dDijk}{ \hat\partial_o D_{kij}  = -2\alpha
\partial_k K_{ij}-2K_{ij}\partial_k\alpha,}
where $K=\gamma^{mn} K_{mn}$ is the trace of the extrinsic curvature, $\nabla_i$ is the covariant derivative operator associated with the three-metric, $R_{ij}$ is the Ricci tensor associated with the three-metric, and $\hat\partial_o=\partial_t-\pounds_\beta$, with $\pounds_\beta$ the Lie derivative along the shift vector $\beta^i$. 

The Lie derivatives of the dynamical variables are 
\labeq{Lie_gamma}{\pounds_{\bf\beta}\gamma_{ij}=\nabla_i\beta_j + \nabla_j \beta_i,}
\labeq{Lie_Kij}{\pounds_{\bf\beta} K_{ij}= (\nabla_i\beta^m) K_{mj} + (\nabla_j\beta^m) K_{mi} + \beta^m \nabla_m K_{ij},}
and
\labeq{Lie_Dkij}{\begin{split}
\pounds_{\bf\beta} D_{kij}=&\ \beta^m\partial_m D_{kij}+D_{mij}\partial_k\beta^m \\
			& \ +2D_{km(i}\partial_{j)}\beta^m+2\gamma_{m(i}\partial_{j)}\partial_k\beta^m.
\end{split}
}

The subset of constraint equations is
\labeq{ADM-H}{
               {\cal H}\equiv\quad  R + K^2 - K_{mn} K^{mn} = 0,
}
\labeq{ADM-M}{
              {\cal M}_i\equiv\quad  \nabla_m K^{m}{}_i - \nabla_i K = 0, \quad
              i=1,2,3,
}
\labeq{1stADM-D}{ {\cal
C}_{kij}\equiv\partial_k\gamma_{ij}-D_{kij}=0.}
where $R$ is the trace of the three-Ricci tensor. ${\cal H}$ is  the Hamiltonian constraint, ${\cal M}_i$ the momentum constraints and ${\cal C}_{kij}$ are new constraints due to the introduction of $D_{kij}$.  The equations as presented above do not include matter terms, because this work focuses on vacuum solutions of the Einstein equations. 

A final set of twelve constraints can be derived by virtue of \eqref{D-defin}, via taking a
spatial derivative and keeping in mind that partial derivatives
commute. Then $D_{kij}$ must also satisfy
\labeq{deriv-D}{{\cal C}_{klij}\equiv\partial_k D_{lij}-\partial_l
D_{kij}=0.}
Constraints \eqref{deriv-D} relate to \eqref{1stADM-D} via the following equation 
\labeq{deriv-D2}{{\cal C}_{klij}=\partial_l
{\cal C}_{kij}-\partial_k {\cal C}_{lij}.}

It is important to note here that in equations \eqref{ADM-gamma}-\eqref{ADM-M}, all spatial partial derivatives of the three-metric are replaced by the corresponding $D_{kij}$ variable. However, in doing so an ambiguity arises due to the fact that there is not a single way to replace the mixed second-order derivatives of the three-metric by first order derivatives of the $D_{kij}$ variables. For example the derivative $\partial_1 \partial_2 \gamma_{ij}$ can be replaced by either $\partial_1 D_{2ij}$ or $\partial_2 D_{1ij}$. There is no rule or recipe on how to map those mixed derivatives of the standard ADM formulation to the first-order derivatives of the $D_{kij}$ variables, when deriving the 1st order ADM formulation. Hence, one has to make a choice to lift this ambiguity. No matter what that choice is, in the continuum limit where $C_{kij}=0$, there is no difference between the numerous final forms the first-order ADM formulation may obtain. However, in the discrete limit where the satisfaction of the $C_{kij}$ constraints is at best approximate, there may be some difference. 

In this work the aforementioned ambiguity is lifted by choosing the most natural way in replacing the mixed derivatives of the three-metric, i.e., they are written in as most symmetric way as possible. This is done by following the suggestion of \cite{KST}, according to which 
\labeq{deriv}{\partial_k \partial_l g_{ij}=\partial_{(k} D_{l)ij}.}

This last modification is equivalent to arbitrarily replacing partial derivatives of $\gamma_{ij}$ by $D_{kij}$ in the spatial Ricci tensor $R_{ij}$, and then recalculating the Ricci tensor as follows
\labeq{Riccimodified}{\hat R_{ij}= R_{ij}+\frac{1}{4}\gamma^{ls}({\cal C}_{iljs}+{\cal C}_{jlis}-{\cal C}_{ijls}),}
where $\hat R_{ij}$ stands for the recalculated Ricci tensor. It is noted here, that it seems that 
the last term in the parenthesis makes the Ricci tensor asymmetric under the interchange of its indices, because that term satisfies ${\cal C}_{ijkl}=-{\cal C}_{jikl}$. However, a straightforward calculation shows that the addition of this term is the only way that the Ricci tensor can be cast into a manifestly symmetric object,  in the context of the 1st order ADM formulation. 

Note that a more general Ricci tensor $\tilde R_{ij}$ can be calculated by adding to the RHS of \eqref{Riccimodified} extra terms as follows
\labeq{Riccimodified2}{\tilde R_{ij}= \hat R_{ij}+\psi\gamma^{ab}{\cal C}_{a(ij)b}.} 
This modification is redundant for system 1 because the system is well-posed and has constraint damping when $\psi=0$, see section~\ref{system_1}. However, $\psi\neq 0$ is essential for the well-posedness of system 2, see section~\ref{system_2}.

%%%%%%%%%%%%%%%%%%%%%%%%%%%%%%%%%%%%%%%%%%%%%%
\subsection{Evolution of the constraints of 1st-order ADM}
%%%%%%%%%%%%%%%%%%%%%%%%%%%%%%%%%%%%%%%%%%%%%%

By taking the time-derivative, $\dt$, of the constraint equations of the first order ADM formulation one can
derive the evolution equations for those constraints by replacing the time derivatives of the dynamical variables through the evolution equations and then eliminate the appropriate combinations of spatial derivatives of the dynamical variables via the constraint equations.  After some tedious algebra, one finds that the evolution of the constraints, without using \eqref{Riccimodified}, is given by 
\labeq{tCkij}{\dt {\cal C}_{kij}=0,}
\labeq{tCklij}{\begin{split}
\dt  {\cal C}_{klij}= 0,
\end{split}}
\labeq{tH}{\begin{split} 
\dt  {\cal H}= & 
2\alpha K {\cal H}-2\alpha \gamma^{ks}\partial_k{\cal
M}_s -4{\cal M}_n\gamma^{nm}\partial_m \alpha \\
& +2\alpha
\gamma^{mn}\Gamma^s_{mn}{\cal
M}_s +H^{mns}{\cal C}_{mns},
%\\ 
%& (\partial_k\alpha)\gamma^{km}(K\gamma^{sa}{\cal C}_{sam}- 
% 2K^b{}_m\gamma^{sa}{\cal C}_{sam}+K^{ab}{\cal C}_{mab}) \\
%& +\alpha (\partial_i K_{sj})(2\gamma^{ka}\gamma^{sb}\gamma^{ij}+\gamma^{sa}\gamma^{jb}\gamma^{ki}-
%   2\gamma^{ks}\gamma^{ia}\gamma^{jb}-\gamma^{ka}\gamma^{ib}\gamma^{sj}) {\cal C}_{kab} \\
%&+2\alpha(\gamma^{ks}\gamma^{ma}\gamma^{nb}-\gamma^{ka}\gamma^{sb}\gamma^{mn})\Gamma^\ell{}_{mn} K_{\ell s} {\cal C}_{kab}
\end{split}
 }
\labeq{tM}{\begin{split}
\dt {\cal M}_i=&\ -\half\alpha\partial_i {\cal H}+\half\alpha\gamma^{mn}\gamma^{\ell s}\partial_\ell{\cal C}_{misn} -{\cal H}\partial_i \alpha \\
          &\ +\alpha K{\cal M}_i  +\ ^{(1)}M^{msn}{\cal C}_{imsn}  +\ ^{(2)}{M}_i{}^{msn\ell}{\cal C}_{msn\ell} \\
			&+\ ^{(3)}M^{mn}{\cal C}_{imn}  +\ ^{(4)}M_i{}^{msn}{\cal C}_{msn},
%+\beta^l \partial_l{\cal M}_i \\ 
%& -\beta^l\gamma^{nm}D_{iml}{\cal
%M}_n+(\partial_i\beta_l)\gamma^{lk}{\cal M}_k
\end{split}
 }
where $H^{mns}, \ ^{(1)}M^{msn}, \ ^{(2)}{M}_i{}^{msn\ell}, \ ^{(3)}M^{mn}, \ ^{(4)}M_i{}^{msn}$ are functions of the dynamical variables and their spatial derivatives, as well as of the lapse function and its spatial derivatives. Their exact forms are long and they are not presented here for brevity, but the full equations are available by the author upon request. 

The Lie derivatives  of the constraints along the shift vector are
\labeq{Lie_Ckij}{\pounds_{\bf\beta} {\cal C}_{kij}=0,}
\labeq{Lie_Cklij}{\pounds_{\bf\beta} {\cal C}_{klij}=\beta^a\partial_a{\cal
C}_{klij}+2(\partial_{[l}\beta^a)C_{k]aij} 
+2C_{kla(j}(\partial_{j)}\beta^a),}
\labeq{Lie_H}{\pounds_{\bf\beta}{\cal H}=\beta^m\partial_m H,}
\labeq{Lie_Mi}{\pounds_{\bf\beta}{\cal M}_i=\beta^m\partial_m M_i+M_m\partial_i\beta^m.}
From equations \eqref{tCkij}-\eqref{tM} it is expected that if the constraints are satisfied on 
the initial time-slice, they should be satisfied throughout the entire 
evolution. 

It is noted here, that if equation \eqref{Riccimodified} is used for the Ricci tensor, it is straightforward to show that the Ricci scalar and hence the Hamiltonian constraint remain unaffected. However, since the evolution equations of the extrinsic curvature are modified, the resulting expressions of the evolution of the momentum and Hamiltonian constraints change. Their exact expressions are not presented here, but their derivation is straightforward, if one uses equations \eqref{tCkij_dyn}-\eqref{M1_4} of appendix \ref{appA}. 

%Note that in \eqref{tCkij}-\eqref{tM} constraints ${\cal C}_{klij}$ are considered independent of ${\cal C}_{kij}$. The evolution of ${\cal C}_{klij}$ is in principle 
%different than that of ${\cal C}_{kij}$ and it is differently affected by adding terms to the RHS of the first-order ADM formulation. In addition to that, ${\cal C}_{klij}%$ can change the stability properties of the evolution equations as it will be shown in subsequent sections. It is also important to note that if  ${\cal C}_{kij}$ are 
%satisfied then ${\cal C}_{klij}$ must be satisfied as well, but the reverse is not true. Moreover, violation of the ${\cal C}_{kij}$ constraints may lead to violation of 
%the ${\cal C}_{klij}$. For all these reasons, constraints \eqref{deriv-D} will be treated as independent throughout this work.  

The principal part of \eqref{tCkij}-\eqref{tM}, has only two non-zero eigenvalues, which propagate with the fundamental speed $\pm\alpha$, when $\beta^i=0$. All remaining eigenvalues are zero (static modes). Static modes are present in the ADM constraint evolution system, even if $\beta^i\neq0$ because the time derivative of ${\cal C}_{kij}$ is zero independently of the shift vector.  

This is where the argument in \cite{Alcubierre} fits in, to explain why several formulations of GR blow up when integrated numerically. Just like the ADM constraint propagation equations, the evolution of the constraints with other formulations, such as KST, have a set of $0$ eigenvalues. This implies that when constraint violating perturbations arise during numerical integration of the evolution equations, these perturbations remain in the system because they cannot propagate off the computational domain. As a result, continuous perturbations by numerical errors lead to accumulation of constraint violating noise which eventually terminates the computations. 

Note, however, that the reason why the ADM formulation has not been successful in multi-dimensional simulations is because the evolution system is ill-posed in more than one dimensions.

This section is concluded by noting that since the ADM constraint equations have to be satisfied at every single point of a spacelike hypersurface, not only the constraints themselves must be zero, but also their
spatial partial derivatives of any order have to be zero, i.e.,
\labeq{con-deriv}{\begin{split}
 \partial_s{}^\nu {\cal C}_{ijk} = & 0,  \\
 \partial_s{}^\nu {\cal C}_{klij} = & 0,  \\
 \partial_s{}^\nu {\cal H}= & 0,  \\
 \partial_s{}^\nu {\cal M}_i = & 0.  \\
\end{split}
}  This means that one is free to add to the
RHS of the ADM equations not only combinations
of the constraints themselves, but also combinations of their
spatial derivatives. This property, will be used in subsequent sections to obtain parabolized formulations of general relativity.

%%%%%%%%%%%%%%%%%%%%%%%%%%%%%%%%%%%%%%%%%%%%%%%%%%%%%%%%%%%%%%%%%%%%%%%%%%%%%%%%%%%%%%%%%%%%
\section{Parabolization of the 1st-order ADM formulation \label{system_1}}
%%%%%%%%%%%%%%%%%%%%%%%%%%%%%%%%%%%%%%%%%%%%%%%%%%%%%%%%%%%%%%%%%%%%%%%%%%%%%%%%%%%%%%%%%%%%

In this section equations \eqref{ADM-gamma} to \eqref{dDijk} are modified in such a way as to have as many parabolic modes as possible. Parabolic modes are  modes which decay exponentially with time at a rate $\exp{(-\epsilon\kappa^2t)}$, with $\epsilon>0$ a damping parameter and $\kappa^2$ the magnitude squared of the wavevector. With that in mind, the following modification is proposed
\labeq{moddgij}{\partial_t \gamma_{ij}=(ADM)+\alpha\lambda
\gamma^{ab}\partial_b {\cal C}_{aij},
}
\labeq{moddKij}{\partial_t K_{ij}=(ADM)+\phi\alpha \gamma_{ij}\gamma^{ab}\partial_a {\cal M}_b
+\theta \alpha \partial_{(i}{\cal M}_{j)},
}
\labeq{moddDkij}{\partial_t D_{kij}=(ADM)+\epsilon\alpha \gamma^{ab}\partial_a {\cal C}_{bkij} +\xi\alpha\gamma_{ij}\partial_k {\cal H}
+\zeta {\cal C}_{kij}, 
}
where $(ADM)$ denotes the RHS of equations \eqref{ADM-gamma} - \eqref{dDijk}. Equations \eqref{moddgij}-\eqref{moddDkij} are called system 1 in this work.

The added derivatives of the Hamiltonian and momentum constraints are given by rather	 complicated and long expressions and for this reason they are not presented here. 
However, their full expressions are available by the author upon request. It is only noted that in those expressions whenever derivatives of the
3-metric, $\partial_k\gamma_{ij}$, are present they are replaced by the variables $D_{kij}$. This ways one ensures that the resulting system
of PDEs is quasi-linear instead of non-linear.

%Parabolization of the ADM formulation has three desired
%properties: 1) any small amplitude short wavelength constraint
%violating perturbations will be damped very quickly, because of
%the diffusive nature of the added terms. Furthermore, the higher the frequency of the perturbations
%the more efficiently they are damped. 2) The modified equations
%are equivalent to the ADM equations on the surface of
%constraints and 3) the evolution equations of the constraints become second-order
%parabolic equations, independently of the gauge choice and spacetime configuration, turning hence the constraint surface into a local attractor. All these will be soon realized in subsequent sections.

%%%%%%%%%%%%%%%%%%%%%%%%%%%%%%%%%%%%%%%
\subsection{Structure and Well-Posedness of system 1 \label{wp_system_1}}
%%%%%%%%%%%%%%%%%%%%%%%%%%%%%%%%%%%%%%%

Any new formulation of GR intended for numerical simulations is required to admit a well-posed initial value problem. This is because an 
ill-posed formulation cannot lead to long-term and stable numerical simulations of the Einstein Equations. This sub-section 
establishes the well-posedness of system 1 and discusses the structure of its evolution equations. 

In the analysis that follows $\zeta$ is considered to be zero because the results with $\zeta \neq 0$ are
rather complicated. It has been found that, just like in the model equation of section~\ref{model_equation}, $\zeta$ has no impact on the well-posedness of system 1, but it must satisfy $\zeta>0$ for long-term and stable numerical simulations.  

For the study of well-posedness of a given set of evolution equations the zeroth-order terms are omitted  and only the higher-order terms have to be considered. For the family of gauges \eqref{algebraic_lapse}  the highest order operator of system 1 is of second-order and according to \cite{PDE_textbook} it has to be analyzed first.  

The second-order part is given by the following equations
\labeq{pp_moddgij}{
\partial_t \gamma_{ij}\approx\alpha\lambda\gamma^{mn}\partial_m\partial_n\gamma_{ij},
}
\labeq{pp_moddKij}{\begin{split}
\partial_t K_{ij}\approx& \ \alpha\phi\gamma_{ij}\gamma^{mn}\gamma^{ks}(\partial_m\partial_k K_{ns}-\partial_k\partial_s K_{mn})+ \\
& \ \half\alpha\theta\gamma^{mn}(\partial_m\partial_{i}K_{jn}+\partial_m\partial_{j}K_{in}-2\partial_i\partial_j K_{mn}),
\end{split}
} 
\labeq{pp_modDkij}{\begin{split}
\partial_t D_{kij}\approx & \ \alpha\gamma^{\ell s}\gamma^{mn}(\partial_k\partial_\ell D_{msn}-\partial_k\partial_\ell D_{smn})+ \\
					& \ \alpha\epsilon\gamma^{mn}(\partial_m \partial_n D_{kij}-\partial_m \partial_k D_{nij}),
\end{split}
} 
where the symbol ``$\approx$" means equal to the second-order part. It can be readily seen that the evolution equations of $\gamma_{ij}, \ K_{ij}$ and $D_{kij}$ decouple to second order. This implies that one can study each set of equations separately. 

The evolution equations for the three-metric \eqref{pp_moddgij} are quasi-linear second-order parabolic, if $\lambda>0$. All six eigenvalues of $P(i \kappa)$ are equal to 
 \labeq{omegg}{\omega_i=\ -\alpha\lambda \kappa^2,\ \ \ i=1,\ldots 6}
and for $\lambda>0$ condition \eqref{parabolic_condition} is satisfied. 

The set of eigenvalues of $P(i \kappa)$ of equation \eqref{pp_moddKij} is 
\labeq{omegK}{\begin{split} \omega_i= &\ -\half\alpha\theta\kappa^2 \ \ \ i=7,8, \\
						\omega_9=&\ 2\alpha\phi\kappa^2, \\
						\omega_i=& \ 0, \ \ \ i=10,\dots,12.
\end{split}
} 
Therefore, $K_{ij}$ does not evolve according to parabolic equations because condition \eqref{parabolic_condition} cannot be satisfied due to the existence of zero eigenvalues. 

An analogous analysis of equations \eqref{pp_modDkij} yields the following eigenvalues 
\labeq{omegD}{\begin{split}
\omega_i= &\ -\alpha\epsilon\kappa^2, \ \ \ i=13,\ldots 24, \\
\omega_{25}=&\ 2\alpha\xi\kappa^2, \\
\omega_i = & \ 0, \ i=26, \ldots 30.
\end{split}
}
Thus, condition \eqref{parabolic_condition}  cannot be satisfied and $D_{kij}$ do not evolve according to parabolic equations either.

Therefore, system 1 is not parabolic and thus theorem 1 cannot be used to prove well-posedness. However, the eigenvalues of the Fourier transformed second-order part of the evolution equations constitute a diagnostic tool for well-posedness and in order for system 1 to be well-posed, the following conditions are necessary
\labeq{pp_cond}{
\lambda>,\theta>0,\phi<0,\xi<0,\epsilon>0.}

Before a sufficient proof of well-posedness is given, note that in appendix~\ref{app_B} it is shown that if the Fourier transform $P(i\kappa)$ of the principal part of a quasi-linear spatial operator $P(\partial_x)$ 
\begin{enumerate}
\item[a)] satisfies the Petrovskii condition~\cite{Gustafsson}, i.e., for all $\kappa$ the eigenvalues $\omega$ of $P(i\kappa)$ satisfy $Re(\omega)\leq c$, where c $\in {\mathbf R}$, 
\item[b)] it has a complete set of eigenvectors and 
\item[c)] the eigenvectors can be chosen as analytic functions for $\kappa\neq0$ 
\end{enumerate}
then equation $\partial_t u=P(\partial_x)u$ admits a well-posed Cauchy problem.

For general second order systems which are not parabolic, the first order terms have to be taken into consideration in the definition of the principal part \cite{PDE_textbook}. Hence, one has to consider both second and first order terms in the principal part of system 1. 

For the family of gauges \eqref{algebraic_lapse} with zero shift vector, the Fourier transformed principal part of system 1 has 28 non-zero eigenvalues, 22 of which have a non-zero real part and the remaining 6 are purely imaginary. Due to the Petrovskii condition, only the real part of the eigenvalues is important for establishing well-posedness. It turns out that the Petrovski condition can be satisfied only when
\labeq{xiphi}{\xi=\phi.}
 If \eqref{xiphi} is true, the real parts of the 22 complex eigenvalues are equal to the 22 non-zero eigenvalues \eqref{omegg}-\eqref{omegD} of the second-order part of system 1. Hence, if both \eqref{pp_cond} and \eqref{xiphi}  hold true,
the Petrovskii condition is satisfied with $c=0$ because the real part of all eigenvalues is strictly non-positive. 

The Fourier transform of the principal part of system 1 is found to have a complete set of eigenvectors, if  one of the following conditions holds true
\labeq{cond1a}{
\theta\neq 2\epsilon, \ \lambda\neq\epsilon
}
or
\labeq{cond1b}{
\theta\neq 2\epsilon, \ \lambda= \epsilon=-2\xi=-2\phi
}

Finally, a calculation of the eigenvectors of $P(i\kappa)$ of system 1 shows that the eigenvectors can be chosen as analytic functions for $\kappa\neq 0$. In particular, their components can be chosen as sums and products of polynomials of $\kappa_i$ and polynomials of $\kappa$. 

Therefore, if \eqref{pp_cond} and \eqref{xiphi} are satisfied and one of equations \eqref{cond1a} or \eqref{cond1b} is true, then conditions a), b) and c) above are met and system 1 admits a well-posed initial value problem. This holds true, if a non-zero fixed shift vector is assumed. 
    
Note here, that the presence of the lapse function as a multiplicative factor in the added terms of system 1 is not necessary for well-posedness and omitting it may prove useful. This is because  the damping rate of the constraints is of the form $\exp(-\alpha\lambda\kappa^2 t)$, and hence when the lapse function is small, the damping time-scale may become large. Examples where the lapse function attains small values are black hole spacetimes. It is well-known that $\alpha$ becomes small near the singularities. A way to overcome the difficulty of inefficient damping, when the lapse function is small, is to allow large values for $\lambda$. However, this could result in numerical complications (see section \ref{parameters}). Therefore, it may eventually be better to remove the lapse function from the damping terms because this results in a damping rate of the form $\exp(-\lambda\kappa^2t)$. 

Another important subject is the structure of the evolution equations of system 1. In addition to the complex eigevalues, the principal part of system 1 has 8 eigenvalues with zero real parts. These are a subset of the eigenvalues of the principal part of the ADM equations and are: $0$ with multiplicity two, $\pm i\alpha\kappa$ each with multiplicity two and $\pm i\alpha\sqrt{2A}\kappa$. Note that, if a non-zero shift vector is considered, the two zero eigenvalues become $\beta^i \kappa_i$. These 8 modes correspond (as is illustrated in section \ref{flat_space_stability}) to constraint-satisfying degrees of freedom, which are the four gauge modes and the four physical modes. In \cite{Paschalidis07} it was shown that for gauge \eqref{algebraic_lapse} the constraint satisfying modes are described by strongly hyperbolic equations. Thus system 1 has a set of hyperbolic modes and a set of parabolic modes and for this reason it may be classified as a mixed hyperbolic - second-order parabolic (MHSP) set of equations. However, notice that it does not really fulfill the requirements of definition 2.

The goal of maximazing the number of non-zero eigenvalues has been achieved by system 1. All degrees of freedom now evolve according to a principal part. This, according to \cite{Alcubierre}, may be an advantage for numerical applications. 

In addition, something even more important has been achieved: The parabolic modes correspond to constraint violating perturbations, which are now getting damped. This can be more easily realized, if one considers a solution to the equations and apply small amplitude constraint satisfying and small amplitude constraint violating perturbations separately. If constraint satisfying perturbations are imposed on the system, then an exact cancellation of the added terms of system 1 occurs and the system reduces to that of ADM. A thorough study of the evolution of constraint satisfying perturbations has been carried out in \cite{Paschalidis07} and hence this is not the subject of this work. If however, one imposes constraint violating perturbations on system 1 then the parabolic terms are turned on. As a result, the constraint violating perturbations and in particular the short-wavelength ones, have no other choice, but to decay exponentially with time. This will also be realized by the structure of the evolution equations of the constraints in the following section.

%A way to overcome this difficulty is to assign to $\lambda$ a spatial dependence. However, it turns out that it is not important for well-posedness to have the lapse function multiply the added terms of system 2.  

%%%%%%%%%%%%%%%%%%%%%%%%%%%%%%%%%%%%%%%%%%%%%%%%%%%%%
\subsection{Evolution of the Constraints with System 1 \label{paraADM_con_evol}}
%%%%%%%%%%%%%%%%%%%%%%%%%%%%%%%%%%%%%%%%%%%%%%%%%%%%%

By virtue of system 1
 and equations \eqref{tCkij_dyn}-\eqref{M1_4}, it can be shown that the constraint evolution equations with system 1 are
\labeq{tCkijparabolic}{\begin{split}
\dt {\cal C}_{kij}=& \ \alpha\lambda\gamma^{ab}\partial_a\partial_b {\cal C}_{kij}-\alpha\zeta {\cal C}_{kij}+\alpha(\lambda-\epsilon)\gamma^{ab}\partial_a{\cal C}_{bkij}  \\ 
& \ -\alpha\xi\gamma_{ij}\partial_k {\cal H}+\alpha\lambda(\partial_k\gamma^{ab})\partial_a {\cal C}_{bij} \\
&\ +(\partial_k\alpha)\lambda\gamma^{ab}{\cal C}_{bij},
\end{split}}
\labeq{tCklijparabolic}{\begin{split}
\dt {\cal C}_{klij}=& \ \alpha\epsilon\gamma^{ab}\partial_a\partial_b {\cal C}_{klij}-\zeta\alpha {\cal C}_{klij}  \\
			& \ + 2(\partial_{[k}\alpha)\big[\epsilon\gamma^{ab}\partial_a{\cal C}_{bl]ij}+\zeta C_{l]ij}+\xi\gamma_{ij}\partial_l{\cal H}\big] \\
				& \ +2\alpha\xi(\partial_{[k}\gamma_{ij})\partial_{l]}{\cal H}+2\epsilon\alpha(\partial_{[k}\gamma^{ab}){\cal C}_{bl]ij},
\end{split}
 }
\labeq{tHparabolic}{\dt {\cal H}\simeq-2\xi\alpha\gamma^{ab}\partial_a\partial_b {\cal H}-\alpha\epsilon\gamma^{si}\gamma^{kj}\gamma^{al}
\partial_s\partial_a{\cal C}_{klij},
}
\labeq{tMiparabolic}{\dt {\cal M}_i\simeq \half\alpha\theta\gamma^{ab}\partial_a\partial_b {\cal M}_i
				-\half(\theta+4\phi)\alpha\gamma^{ab}\partial_i\partial_a {\cal M}_b,
}
where the exact evolution equations of ${\cal C}_{kij}$ and ${\cal C}_{klij}$ have been presented, but only the principal parts of the evolution equation of ${\cal H}$ and ${\cal M}_{i}$ have been given (``$\simeq$" means equal to the principal part). 

The eigenvalues of the Fourier transform of the principal part of system \eqref{tCkijparabolic}-\eqref{tMiparabolic} are all non-zero and they equal the non-zero eigenvalues of the second-order part of system 1 \eqref{omegg}-\eqref{omegD}. Thus, equations \eqref{tCkijparabolic}-\eqref{tMiparabolic} are second-order parabolic, when condition \eqref{pp_cond} is satisfied and hence they admit a well-posed Cauchy problem. 

The parabolic structure of the evolution equations of the constraints implies that the constraints are damped with time. In particular, the shortest the wavelength of perturbations which violate the constraints,  the faster those perturbations are damped because the damping rate is of the form $\exp(-\alpha\lambda \kappa^2t)$. The new evolution equations of the constraints suggest that the surface of constraints is a local attractor.  

Another important property of system 1 is that the constraint violatiions are smoothed with time. This is due to the wavelength-dependent nature of the damping timescale. 

Furthermore, the structure and hence the smoothing and damping properties of equations \eqref{tCkijparabolic}-\eqref{tMiparabolic} is independent of the gauge condition or the spacetime solution, since no specific gauge or spacetime geometry was assumed in the derivation of the new constraint propagation equations.

Finally, equations \eqref{tCkijparabolic}, \eqref{tCklijparabolic} and \eqref{tMiparabolic}  are completely decoupled to highest-order from the rest of the system and are manifestly second-order parabolic, when \eqref{pp_cond} is satisfied. However, \eqref{tHparabolic} is coupled to second-order to \eqref{tCklijparabolic}. 

At this point, one may wonder if the aforementioned discussion on damping and well-posedness is relevant when equation \eqref{deriv-D2} is enforced at the RHS of \eqref{tHparabolic}. This is so, because this enforcement turns \eqref{tHparabolic} into third-order. Even in this case, it is straightforward to show that the eigenvalues remain the same and that the constraint propagation system remains well-posed. The proof of well-posedness of the constraint evolution system, with \eqref{deriv-D2}  enforced at the RHS of \eqref{tHparabolic}, is exactly the same as the proof of well-posedness for parabolic systems presented in \cite{PDE_textbook}. For this reason it is not repeated here. Therefore, the properties of the evolution equations of the constraints are independent of whether \eqref{deriv-D2} is enforced or not. This is to be anticipated because all forms the constraint propagation equations \eqref{tCkijparabolic}-\eqref{tMiparabolic} may obtain by use of \eqref{deriv-D2}  are equivalent with each other and physical conclusions do not depend on the 
form of the equations.

%%%%%%%%%%%%%%%%%%%%%%%%%%%%%%%%%%%%%%%%%%%%%%%%%%%%%%%%%%%%%%%%%%%%%%%%%%%%%%%%%%%%%%%%%%%%
\section{Parabolic Modifications to the KST formulation \label{system_2}}
%%%%%%%%%%%%%%%%%%%%%%%%%%%%%%%%%%%%%%%%%%%%%%%%%%%%%%%%%%%%%%%%%%%%%%%%%%%%%%%%%%%%%%%%%%%%

In this section system 2 is introduced. This system is based on the KST formulation as presented in \cite{KST}. Thus, before going into the details of system 2,  first  the KST formulation is reviewed and the original gauge choice of the KST system is extended to the family \eqref{algebraic_lapse}.

%%%%%%%%%%%%%%%%%%%%%%%%%%%%%%%%%%%%%%%%%%%%%%
\subsection{The KST formulation \label{KST_formulation}}
%%%%%%%%%%%%%%%%%%%%%%%%%%%%%%%%%%%%%%%%%%%%%%

The KST modification to the first-order ADM formulation, as
originally introduced, is a parameterized family of strongly hyperbolic formulations. 
%Exact definitions and theorems pertaining to quasi-linear hyperbolic systems of equations
%can be found in \cite{PDE_textbook}.

The KST modification is
\labeq{KST}{\begin{split}
\partial_t K_{ij}= & (ADM)+\rho \alpha \gamma_{ij}\mathcal{H}+\psi \alpha \gamma^{ab}\mathcal{C}_{a(ij)b}, \\
\partial_t D_{kij}=& (ADM)+\eta \alpha \gamma_{k(i}\mathcal{M}_{j)}+\chi
\alpha \gamma_{ij}\mathcal{M}_k,
\end{split}
}

In addition to these modifications to the RHS of the ADM system, Kidder, Scheel and Teukolsky used a densitized lapse \eqref{dense-lapse}
and a fixed shift vector $\beta^i$.

In this work the lapse condition is extended to the algebraic family \eqref{algebraic_lapse}. The resulting KST system is essentially the same as the original one, but with one difference: $\sigma$ must be replaced by the quantity $A$ defined in equation \eqref{A}. For illustration, the principal part of the second-order covariant derivative of the lapse is
\labeq{dKij2lapse}{\nabla_i\nabla_j \alpha \simeq \half \alpha A
\gamma^{kl}(\partial_j D_{ikl}+\partial_i D_{jkl})%+ \alpha R_{ij}^{1}
                   .}
%
%where the terms $\rho \alpha \gamma_{ij}\mathcal{H}+\psi \alpha \gamma^{ab}\mathcal{C}_{a(ij)b}$ have been omitted for clarity and brevity, and
%where 
%\labeq{Ricci_principal}{\begin{split}
%R_{ij}^1\simeq & \ \ \half\gamma^{mn}(2\partial_mD_{(ij)n}-\partial_mD_{nij})-\half\gamma^{mn}\partial_iD_{jmn} \\
%		& \ +\frac{1}{4}\gamma^{ls}({\cal C}_{iljs}+{\cal C}_{jlis}-{\cal C}_{ijls})
%\end{split}
%}
%
%is the principal part of the Ricci tensor. 
The RHS of \eqref{dKij2lapse} is exactly the same as the RHS it has in the original KST system \cite{KST}, but with $\sigma$ replaced by $A$.

The 30 characteristic speeds of the  principal part of the KST formulation in the generalized gauge \eqref{algebraic_lapse}  are equal to $\{0,\pm \alpha, \pm c_1 \alpha,\pm c_2 \alpha,\pm c_3 \alpha\}$. Eighteen of the speeds are 0. Speeds  $c_1$, $c_2$, $c_3$ are different than those of the original KST formulation and are given by
\labeq{speeds}{\begin{split} c_1^2= & \ 2A, \\
c_2^2= & \ \frac{1}{8}( \eta-4 \eta A-2\chi-12
A \chi-6\eta \psi), \\
c_3^2= & \ \frac{1}{2}( 2+8\rho-\eta-4\rho
\eta+2\chi+8\rho\chi-2\eta \psi),
 \end{split}}
where the squares of the speeds is shown. For the system to be well-posed the RHS of equation \eqref{speeds} must be positive.
% Note that $c_1$ corresponds to a gauge wave, whereas $c_2$ and $c_3$ correspond to constraint violating waves.

The analysis of well-posedness of the KST system in the extended gauge is the same as that of the original one. Hence, if the RHS of equation \eqref{speeds} is positive, the new evolution system is strongly hyperbolic unless one of the following conditions is satisfied 
\labeq{strong_hyp}{\begin{split}
 c_i= 0, \ i= & 1,3 \\
 c_1= c_3\neq & 1, \\
 c_1=c_3=1 \neq & c_2.
 \end{split}}

%Note that the only choice that
%can result in physical speeds for the gauge waves is a harmonic
%lapse, i.e., $\alpha=Q \gamma^{1/2}$. However, this does not limit
%one's choices, since the gauge degrees of freedom may propagate
%faster than light. 

Ensuring globally that conditions \eqref{strong_hyp} will not be satisfied is not as straightforward anymore due to the spatial dependence of $A$. However, there is a way around this difficulty. Notice that if speeds $c_2, c_3$ are set to unity, the strong hyperbolicity conditions are met. This can be done irrespectively of whether densitized lapse or ``$1+\log$'' slicing is chosen, by requiring that the following conditions be satisfied
\labeq{c2_c3_1}{
\begin{split}
\eta+3\chi= 0, \ \ \eta-2\chi-6\eta\psi= & 8, \\
2-\eta+8\rho-4\eta\rho +2\chi+8\rho\chi-2\eta\psi= & 2.
\end{split}
}
Solving these equations parametrically one obtains
\labeq{c2_c3_1b}{
\eta=-3\chi, \ \ \ \psi=\frac{5\chi+8}{18\chi}, \ \ \ \rho=-\frac{1}{3}, \ \ \  \chi\neq \{0,-8/5\}
}
or 
\labeq{c2_c3_1c}{\eta=\frac{6}{5}, \ \ \ \psi=-\frac{5}{6}, \ \ \ \chi=-\frac{2}{5}, \ \ \  \rho\neq 0.
}
The choice $\chi=-8/5$ in \eqref{c2_c3_1b} renders the system weakly hyperbolic and hence ill-posed. The solutions above coincide for the following values
\labeq{c2_c3_1d}{
\eta=\frac{6}{5}, \ \ \ \psi=-\frac{5}{6}, \ \ \ \chi=-\frac{2}{5}, \ \ \  \rho= -\frac{1}{3}.
}
%A flat space stability analysis of the KST formulation (see section \ref{flat_space_KST}) shows that if the parameters are chosen according to equation \eqref{c2_c3_1d},  the number of flat space modes, which violate the Hamiltonian constraint, is smaller than any other choice. 

The generalization of the gauge constitutes the first extension of the original KST formulation. 

% The terms added 
%to the RHS of the ADM evolution equations aim at controlling the growth of the constraint violating modes under free evolution. This can again be 
%understood by considering constraint satisfying perturbations and constraint violating perturbations separately. If
%constraint satisfying perturbations are imposed on the system, then the added terms are identically zero and the behavior 
%of those modes follows the analysis presented in \cite{Paschalidis07}. However, if constraint violating modes are imposed the added terms are switched on and they dictate to those perturbations how to evolve, which in the KST case they force them to propagate as waves. 

%%%%%%%%%%%%%%%%%%%%%%%%%%%%%%%%%%%%%%%%%%%%%%%%%%%%%%%
\subsection{Evolution of the constraints of the KST modification to ADM}
%%%%%%%%%%%%%%%%%%%%%%%%%%%%%%%%%%%%%%%%%%%%%%%%%%%%%%%

By virtue of equations \eqref{tCkij_dyn}-\eqref{M1_4} , one can show that the evolution of the constraints with the KST formulation becomes
\labeq{tCkijKST}{
\dt {\cal C}_{kij}=-\eta\alpha\gamma_{k(i}{\cal M}_{j)}-\chi\gamma_{ij}{\cal M}_{k},
}
\labeq{tCklijKST}{\dt {\cal C}_{klij}=2\chi\alpha\gamma_{ij}\partial_{[k}{\cal M}_{l}
						+\alpha\eta\big(\gamma_{i[l}\partial_{k]}{\cal M}_{j}+\gamma_{j[l}\partial_{k]}{\cal M}_{i}\big),
}
\labeq{tHKST}{\dt {\cal H}\simeq \alpha(\eta-2\chi-2)\gamma^{mn}\partial_m{\cal M}_{n},
}
\labeq{tMiKST}{\begin{split} \dt {\cal M}_{i}\simeq & \ -\half\alpha(1+4\rho)\partial_i {\cal H}
				+\frac{1}{4}\alpha(1+2\psi)\gamma^{ab}\gamma^{ls}\partial_l {\cal C}_{aisb} \\
			&\ -\frac{1}{4}\alpha(1-2\psi)\gamma^{ab}\gamma^{sk}\partial_k {\cal C}_{saib} \\
			&\ -\frac{1}{4}\alpha(1+2A)\gamma^{ab}\gamma^{sk}\partial_k {\cal C}_{siab},
\end{split}
}
where the exact evolution equations of the ${\cal C}_{kij},\ {\cal C}_{klij}$ constraints are presented, but only the principal part of the evolution of  ${\cal H}, \ {\cal M}_{i}$ is given. The apparent discrepancy in these equations with the equations presented in \cite{KST}, arises because of the difference of the definitions of ${\cal H} \mbox{\ and\ } {\cal C}_{klij}$. In particular, ${\cal H}=2\hat{\cal H}$ and ${\cal C}_{klij}=2\hat{\cal C}_{klij}$,  where $\hat{\cal H}$ and $\hat{\cal C}_{klij}$  denote the definition of those constraints in \cite{KST}.

The characteristic speeds of the Fourier transform of the principal part of equations \eqref{tCkijKST}-\eqref{tMiKST} are  $\{0,\pm c_2\alpha,\pm c_3\alpha\}$, i.e., they are a subset of the characteristic speeds of the evolution equations, as expected. It was shown in \cite{KST} that if the evolution equations of the dynamical variables are strongly hyperbolic, then the constraint propagation equations are also strongly hyperbolic. 

Although this property is really desirable, the KST formulation has a serious disadvantage. Equation \eqref{tCkijKST}, shows that ${\cal C}_{kij}$  evolve according to lower order terms. This implies that ``static" modes correspond to  the principal of \eqref{tCkijKST}. These low order terms of \eqref{tCkijKST} are functions of the momentum constraints. This in turn means that any violations of the momentum constraints will directly lead to violations of ${\cal C}_{kij}$. This is not desirable and therefore damping of  ${\cal C}_{kij}$ should be introduced in the KST system. %This has already been numerically confirmed.

%Simulations presented in \cite{PaschalidisHansen} demonstrate this important disadvantage of the KST system and show how the excitation of ``static modes" by %violations of the momentum constraints can lead to exponential blowups. 

%In addition, the results obtained in \cite{Robert_Owen}, show that the KST system in conjunction with damping of the ${\cal C}_{kij}$ constraints substantially 
%prolongs the simulations of a single black hole. 

%Therefore, there is sufficient evidence which shows that perturbations by  low order terms can lead to exponentially growing error because of the presence of 
%static modes in the KST evolution system, unless constraint damping is introduced at the analytic level of the equations.  

%%%%%%%%%%%%%%%%%%%%%%%%%%%%%%%%%%%%%%%%%%%%
\subsection{Mixed Hyperbolic-Parabolic KST Formulation \label{mod_KST}}
%%%%%%%%%%%%%%%%%%%%%%%%%%%%%%%%%%%%%%%%%%%%

In this section the generalized KST system of section~\ref{KST_formulation} is modified so that the resulting formulation has the desirable properties of strong damping and smoothing of ${\cal C}_{kij}$. This is achieved by the same parabolization strategy that was applied to the scalar wave equation of section~\ref{model_equation}. % and  the ADM formulation~\ref{system_1}.  
In particular, the following modification is suggested
\labeq{Modified_KST}{\begin{split}
\dt \gamma_{ij}= & (KST)+\lambda \alpha \gamma^{ab}\partial_a {\cal C}_{bij}, \\
\dt D_{kij}=& (KST)+\zeta\alpha {\cal C}_{kij},
\end{split}
}
where $(KST)$ stands for the RHS of the extended KST formulation as that is given in section \ref{KST_formulation}. Equations \eqref{Modified_KST} together with the evolution equations of $K_{ij}$ from equations \eqref{KST} are called system 2 throughout this work.

For $\lambda>0$ the evolution equation of the three-metric is a parabolic equation (see section~\ref{sys2_wp}). Notice also the term $\zeta\alpha {\cal C}_{kij}$ which has been added to the evolution of $D_{kij}$. This is important to couple the ${\cal C}_{kij}$ constraint violating modes which arise from the evolution equations of the three-metric, to those which arise from the evolution equations of the $D_{kij}$ variables. It is straightforward to check that if $\lambda=0$, then there still are eighteen zero eigenvalues in the principal part of the formulation, whereas if $\zeta=0$ then there are twelve zero eigenvalues. As in equation \eqref{modified_wave_equation}, if both $\lambda> 0$ and $\zeta> 0$, strong damping of ${\cal C}_{kij}$ occurs. The $\lambda$ term takes care of the short wavelength perturbations, whereas the $\zeta$ term takes care of the long wavelength ones. 

Note that the proposed modification can be easily integrated with the system of redefined variables, which was introduced in \cite{KST}. This is so, because in that system only $K_{ij}$ and $D_{kij}$ were transformed. This may be advantageous, because the redefined system has shown better stability properties in numerical simulations than those of the KST formulation presented in section \ref{KST_formulation}.

%%%%%%%%%%%%%%%%%%%%%%%%%%%%%%%%%%%
\subsubsection{Well-posedness of system 2 \label{sys2_wp}}
%%%%%%%%%%%%%%%%%%%%%%%%%%%%%%%%%%%

To prove well-posedness for system 2, note that the equations can symbolically be written as
\labeq{moddgij2b}{\dt \gamma_{ij}=(KST)+\alpha\lambda
\gamma^{ab}\partial_a\partial_b\gamma_{ij}-\alpha\lambda\gamma^{ab}\partial_bD_{aij},}
\labeq{moddKij2b}{\dt K_{ij}=(KST),}
\labeq{moddDkij2b}{\dt D_{kij}=(KST)+\zeta\alpha(\partial_k \gamma_{ij}-D_{kij}).}
It can be readily realized that equations \eqref{moddgij2b}-\eqref{moddDkij2b} have obtained the form of a MHSP system
\eqref{MixedPH} with $u=\gamma_{ij}$ and $v=\{K_{ij},D_{kij}\}$. The coupling terms are $R_{12}=-\alpha\lambda\gamma^{ab}\partial_aD_{bij}$ and $R_{21}=\zeta\alpha\partial_k \gamma_{ij}$ which are first order. The uncoupled \eqref{moddgij2b} is a second-order  parabolic equation provided that $\lambda>0$  and the uncoupled system of  \eqref{moddKij2b} and \eqref{moddDkij2b} is strongly  hyperbolic.  Therefore, the total set of equations is mixed second-order parabolic - strongly hyperbolic, bearing thus resemblance to the compressible Navier-Stokes equations. According to theorem 2, this system admits a well-posed Cauchy problem. 

The value of $\zeta$ does not play any role in the establishment well-posedness, but it is really important for the stability of numerical simulations. If $\zeta <0$, then at some point exponential explosion of the error of ${\cal C}_{kij}$ will take place. Thus, as it was explained in section~\ref{model_equation},
$\zeta >0$ is required for long-term and stable numerical simulations.

Finally, note is that the lapse function  multiplying the $\lambda$ and $\zeta$ terms in equation \eqref{Modified_KST} may be dropped and the system will still be well-posed. As is discussed in section \ref{parameters}, this may be advantageous in spacetimes, where black holes are involved.

\subsection{Evolution equations of the constraints of the parabolized KST formulation}

The modifications introduced to KST in equation \eqref{Modified_KST}, result into modifications of the evolution equations of the constraints. These modifications can easily be derived by virtue of equations \eqref{tCkij_dyn}-\eqref{M1_4} and the resulting constraint propagation system is

\labeq{tCkijsystem2}{\begin{split}
\dt{\cal C}_{kij}= &\ \ \alpha\lambda \gamma^{ab}\partial_a \partial_b {\cal C}_{kij}-\alpha\zeta C_{kij} \\
				& \ +\alpha\lambda\gamma^{ab}\partial_a {\cal C}_{bkij}+\alpha\lambda(\partial_k\gamma^{ab})\partial_a {\cal C}_{bij} -\\
				& \ -\alpha\eta\gamma_{k(i}{\cal M}_{j)}-\alpha\chi\gamma_{ij}{\cal M}_k.
\end{split}
}
\labeq{tH1-tM1}{\begin{split}
\partial_t H\simeq & \ (\dots)+\alpha\big(\zeta\ ^{(1)}{\cal H}^{abij}+\lambda\gamma^{ab}\ ^{(4)}{\cal H}^{ij}\big)\partial_a{\cal C}_{bij} \\
%\alpha \zeta(\gamma^{in}\gamma^{jm}-\gamma^{ij}\gamma^{mn})\partial_i C_{jmn}
%        +2\alpha(\eta-\chi-1)\gamma^{ij}\partial_i M_j
%}
%
%\labeq{tCklij1}{
\partial_t{\cal C}_{klij}\simeq & \ (\ldots),\\
%\zeta\alpha \partial_{[k}{\cal C}_{l]ij}+
%         2\eta \alpha(\gamma_{i[l}\partial_{k]}M_j+\gamma_{j[l}\partial_{k]}M_i)+\chi\alpha\gamma_{ij}\partial_{[k}M_{l]}
%}
%
%\labeq{tM1}{
\partial_t M_i\simeq & \ (\ldots)
%+\sigma\alpha\gamma^{ab}(\gamma^{mj}\partial_j{\cal C}_{a(mi)b}-
%                            \gamma^{mn}\partial_i{\cal C}_{a(mn)b}
, 
\end{split}}
where $(\ldots)$ stands for the RHS of equations \eqref{tCklijKST}-\eqref{tMiKST} and where  $\ ^{(1)}{\cal H}^{abij}$ and $\ ^{(4)}{\cal H}^{ij}$ are given in appendix~\ref{appA}. In the equations above, the exact RHS has been written for the evolution of ${\cal C}_{kij}$ and only the principal parts for the remaining constraint variables.

Equation \eqref{tCkijsystem2} resembles the form of \eqref{modified_wave_equation} and the evolution of ${\cal C}_{kij}$ is not left at the mercy of the low order terms as in the KST formulation because the principal part is not zero anymore. 

Equations \eqref{tCkijsystem2} and \eqref{tH1-tM1} have obtained the form of MHSP just like the evolution equations of system 2. In particular, equation \eqref{tCkijsystem2} is parabolic, the uncoupled system \eqref{tH1-tM1} is strongly hyperbolic and the coupling terms are of first order. Therefore, the constraint propagation equations with system 2 admit a well-posed initial value problem. In addition, due to the parabolic nature of equation \eqref{tCkijsystem2}, the ${\cal C}_{kij}$ constraints are damped and smoothed with time. 

%Just like in the case of the evolution of the constraints with system 1, by using the proposition of Appendix~\ref{app_B} it can be shown that the well-posedness 
%and damping properties of equations \eqref{tCkijsystem2} and \eqref{tH1-tM1} are independent of whether equation \eqref{deriv-D2} is enforced at the RHS of 
%\eqref{tCkijsystem2}. This is to be anticipated again because all forms of the evolution equations of the constraints are equivalent with each other.  

%%%%%%%%%%%%%%%%%%%%%%%%%%%%%%%%%%%%%%%%%%%%%%%%%%%%%%%%%%%%%%%%%%%%%%%%%%%%%%%%%%%%%%%%%
\section{Stability of Flat Space \label{flat_space_stability}}
%%%%%%%%%%%%%%%%%%%%%%%%%%%%%%%%%%%%%%%%%%%%%%%%%%%%%%%%%%%%%%%%%%%%%%%%%%%%%%%%%%%%%%%%%

In this section the stability of flat space is studied in conjunction with system 1, the first-order ADM and the KST formulations. 

System 2 will not be considered any further in this work so that the theoretical predictions of this section are made more relevant to the numerical simulations shown in \cite{PaschalidisHansen}. A flat space stability analysis along with numerical simulations with system 2 will be considered elsewhere. %The purpose of system 2 is to illustrate how one can apply the parabolization strategy to hyperbolic formulations of the Einstein equations.

The stability analysis is carried out by linearizing the evolution equations and the constraints about flat space. In particular small perturbations are applied on all dynamical variables as follows \mbox{$\gamma_{ij}=\delta_{ij}+\delta\gamma_{ij}$}, $K_{ij}=0+\delta K_{ij}$, $D_{kij}=0+\delta D_{kij}$. The linearized evolution equations show how these perturbations evolve with time, while the linearized constraint equations serve as a tool for the classification of the individual modes in constraint satisfying and constraint violating. Both the evolution and the constraint equations are studied in the Fourier space in order to reduce the analysis to an eigenvalue problem. 

In what follows to make the comparison among the three formulations appropriate the same gauge condition is employed. This gauge condition has zero shift vector and lapse function of the form  $\alpha=\alpha(\gamma)$ with $A=\partial\ln \alpha/\partial\ln \gamma>0$. When linearization is carried out about flat space, ``1+log" slicing cannot be distinguished from densitized lapse, because $A=1$ for both ``1+log" slicing and a densitized lapse with $\sigma=1$. 

In addition to a specific gauge condition, only perturbations along the $x^1$-direction will be considered. The results for any direction are qualitatively the same, but because they are rather complicated, here only a one dimensional analysis is presented. 

The linearized evolution equations are 
\labeq{flat_space_evol1}{%\begin{split}
\partial_t \delta\gamma_{ij}=\ -2\delta K_{ij}+\lambda\partial_x\partial_x\delta\gamma_{ij}-\lambda\partial_x D_{1ij},} %\\
\labeq{flat_space_evol1b}{\begin{split}
\partial_t \delta K_{ij}= 
% (ADM)
					&\ \ \frac{1}{4}[\partial_x\delta D_{ji1}+\delta_{1j}\partial_x \delta D_{kik}+\partial_x \delta D_{ij1} \\
					&\ +\delta_{1i}\partial_x\delta D_{kjk}-(1+2A)\delta_{1i}\partial_x\delta D_{jkk} \\
					&\ -(1+2A)\delta_{1j}\partial_x \delta D_{ikk}
					-2\partial_x \delta D_{1ij}]  \\
%(KST)
					&\ +\rho\delta_{ij}(\partial_x D_{mm1}-\partial_x D_{1mm}) \\
					&\ +\psi(\partial_x D_{(ij)1}-\delta_{1(i}\partial_x D_{a j)a}) \\
%
% PARABOLIZED ADM
%
					&\ -\phi \delta_{ij}(\partial_x \partial_x\delta K_{22}+\partial_x \partial_x\delta K_{33}) \\
					&\ +\half\theta(\delta_{1j}\partial_x \partial_x \delta K_{1i}+\delta_{1i}\partial_x \partial_x \delta K_{1j})\\
					&\ -\theta\delta_{1i}\delta_{1j}(\partial_x \partial_x \delta K_{11} +\partial_x \partial_x \delta K_{22}+
					   \partial_x \partial_x \delta K_{33}), \\
\end{split}
}
\labeq{flat_space_evol2}{\begin{split}
\partial_t \delta D_{1ij} = 
% ADM
				&\ -2\partial_x\delta K_{ij} 
% KST        
				 +\eta \big(\delta_{1(i}\partial_x K_{j)1}-\delta_{1(i}\delta_{j)1}\partial_x K_{mm}\big) \\
				&\ -\chi\delta_{ij}(\partial_x K_{22}+\partial_x K_{33}) \\
%
%PARABOLIZED ADM
& +\xi\delta_{ij}(\partial_x\partial_x \delta D_{m1m}-\partial_x\partial_x \delta D_{1mm}), \\
\partial_t \delta D_{2ij} = 
%KST
				&\ \eta \big(\delta_{2(i}\partial_x K_{j)1}-\delta_{2(i}\delta_{j)1}\partial_x K_{mm}\big) \\				&\ + \chi\delta_{ij}\partial_x K_{12},
% PARABOLIZED ADM
+ \epsilon\partial_x\partial_x\delta D_{2ij}, \\
\partial_t \delta D_{3ij} = 
%KST
				&\ \eta \big(\delta_{3(i}\partial_x K_{j)1}-\delta_{3(i}\delta_{j)1}\partial_x K_{mm}\big) \\				&\ + \chi\delta_{ij}\partial_x K_{13}
% PARABOLIZED ADM
+ \epsilon\partial_x\partial_x \delta D_{3ij}.
\end{split}
}
Equations \eqref{flat_space_evol1}-\eqref{flat_space_evol2} contain all three formulations (namely ADM, KST, system 1) for brevity, because they all have the ADM one as the underlying  formulation. Just by choosing the free parameters one can go from one formulation to another. So, the ADM equations are obtained by setting all free parameters to zero, the KST equations are extracted from \eqref{flat_space_evol1}-\eqref{flat_space_evol2} by setting $\{\lambda=\epsilon=\phi=\theta=\xi=0\}$ and finally system 1 is obtained from \eqref{flat_space_evol1}-\eqref{flat_space_evol2}, by setting $\rho=\psi=\chi=\eta=0$. Note that one could study the system of equations \eqref{flat_space_evol1}-\eqref{flat_space_evol2} as a whole, but this analysis is complicated and will be the subject of a future work.

The linearized constraint equations are
\labeq{Lin_Constraints}{
\begin{split}
\delta {\cal C}_{1ij}\equiv & \  \partial_x \delta \gamma_{ij}-\delta D_{1ij}=0, \\
\delta {\cal C}_{2ij}\equiv & \ -\delta D_{2ij}=0, \\
\delta {\cal C}_{3ij}\equiv & \ -\delta D_{3ij}=0, \\
\delta {\cal H}\equiv &\ \partial_x \delta D_{212}+\partial_x \delta D_{313}-\partial_x \delta D_{122}-\partial_x \delta D_{133}=0, \\
\delta {\cal M}_1 \equiv &\ -\partial_x \delta K_{22}-\partial_x \delta K_{33} =0,\\
\delta {\cal M}_2 \equiv &\ \partial_x \delta K_{12} =0,\\
\delta {\cal M}_3 \equiv &\ \partial_x \delta K_{13} =0.
\end{split}
}

The next step is to  Fourier transform equations \eqref{flat_space_evol1}-\eqref{Lin_Constraints} by requiring a solution of the form
\labeq{Harmonic_pert}{
\begin{split}
\delta \gamma_{ij}= & \tilde\gamma_{ij} e^{i(\kappa x-\omega t)}, \\
\delta K_{ij}= & \tilde K_{ij} e^{i(\kappa x-\omega t)}, \\
\delta D_{kij}= & \tilde D_{kij} e^{i(\kappa x-\omega t)}.
\end{split}
}
The effect of equation \eqref{Harmonic_pert} is the same as replacing a partial spatial derivative, $\partial_x$, with $i\kappa$ and a partial time-derivative, 
$\partial_t$, with $i\omega$.  The Fourier transform of equations \eqref{flat_space_evol1}-\eqref{Lin_Constraints} is presented in appendix~\ref{app_C}.

To study flat space stability in conjunction with the ADM, KST and  parabolized-ADM (system 1) equations, all one has to do is solve the eigenvalue problem  which results  by the Fourier transform of \eqref{flat_space_evol1} and which can be written in the form
\labeq{eig_prob}{M{\bf v}=\omega {\bf v}
}
where the transpose of the column vector ${\bf v}$ is ${\bf v}^T=\{\tilde\gamma_{11},\tilde\gamma_{12},\tilde\gamma_{13},\tilde\gamma_{22},
\tilde\gamma_{23},\tilde\gamma_{33},\tilde K_{11},\tilde K_{12},\tilde K_{13},\tilde K_{22},\tilde K_{23}$, $\tilde K_{33}, \tilde D_{111},\tilde D_{112}, \tilde D_{113},\tilde D_{122},\tilde D_{123},\tilde D_{133},\tilde D_{211},\tilde D_{212},\tilde D_{213}$,
$\tilde D_{222},\tilde D_{223},\tilde D_{233},\tilde D_{311},\tilde D_{312},
\tilde D_{313},\tilde D_{322},\tilde D_{323},\tilde D_{333}\}$ and $M$ is the matrix whose eigenvalues and eigenvectors one has to look for. 

One can check whether these modes violate or satisfy the constraints, via pure inspection of equations \eqref{Fourier_Lin_Constraints}. 
If the components of a given eigenvector of $M$ are such that it makes all constraints vanish, the eigenvector is said to be a constraint-satisfying mode. Otherwise, it is called a constraint-violating mode.

The $30\times 30$ matrix $M$ and its eigenvectors will not be presented in this work for brevity, but they are available  by the author upon request.
What will be presented are its eigenvalues $\omega_i$ or characteristic speeds $v_i=\omega_i/\kappa$, where appropriate. \\

\begin{widetext}
\begin{center}
\begin{table} 
\begin{tabular}{ccccc}\hline\hline
\multicolumn{1}{p{2.8 cm}}{Number of modes\quad} & 
\multicolumn{1}{c}{Speed $v_i$ \quad} & 
\multicolumn{1}{c}{Violates ${\cal C}_{kij}$? \quad} & 
\multicolumn{1}{c}{Violates ${\cal H}$? \quad} & 
\multicolumn{1}{c}{Violates ${\cal M}_{i}$?\quad} \\ \hline\hline
%(1)-(2)
2 		  & 0		  &	 x				         & x                             &              x                    \\  \hline
%(3)-(4)
2		 &  $\pm\sqrt{2A}$  & x	         & x                             &              x                    \\  \hline 
%(5)-(8)
4          &  $\pm	1$ & x			         & x                             &              x                    \\  \hline
%(9)-(10)
2		& $\pm 1$  &  x				&	\checkmark        &      \checkmark      \\ \hline
%(11)-(26)
16       &   0                     &  \checkmark    &         x                   &            x              \\ \hline
\end{tabular}
\caption{Classification of the ADM flat space modes into constraint satisfying and constraint violating. The first column gives the number of  modes. The second column gives the characteristic speed, which corresponds to the modes of the first column. The third, fourth and fifth columns have either a ``x" or a ``\checkmark" as values. A ``x" means that the constraint is not violated by the specific modes, whereas a checkmark ``\checkmark" implies that at least one of the components of the constraint is violated by the specific modes.}	%
\label{table1}
\end{table}
\end{center}
\end{widetext}
\subsection{Flat space stability with the first-order ADM formulation \label{flat_space_ADM}}
%%%%%%%%%%%%%%%%%%%%%%%%%%%%%%%%%%%%%%%%%%%%%%%%%%%%%

%The study of stability of flat space in conjunction with the first-order ADM formulation can be carried out by analyzing the linearized evolution equations. These 
%can be extracted from equations \eqref{Fourier_flat_space_evol1}-\eqref{Fourier_flat_space_evol3}, if $\lambda=\xi=\epsilon=\phi=\theta=0$ and $\rho=\eta=
%\chi=\psi=0$.  
When the ADM equations are employed, $M$ has ten non-zero eigenvalues and twenty-six eigenvectors, if $A\neq 1/2$, whereas it has twenty-four eigenvectors, if $A=1/2$. %The thirty modes of matrix $M$, for $A\neq 1/2$ are given below. 
A classification of these modes into constraint satisfying and constraint violating is presented in table~\ref{table1}.

No matter what the value of $A$ is, there always are eight constraint satisfying modes (first three rows of table~\ref{table1}). In addition to these eight modes, there always are two constraint violating wave modes (fifth row of table~\ref{table1}). 
%These were also expected because the evolution equations of the constraints \eqref{tCkij}-\eqref{tM} contain two wave modes and it is known that the eigenvalues of the constraint propagation equations are a subset of the eigenvalues of the evolution equation of the dynamical variables. 

Note here that the results shown in table~\ref{table1} are valid only for $A\neq1/2$. The case $A=1/2$ has to be considered separately. The results with $A=1/2$, are qualitatively the same, but as already mentioned matrix $M$ has twenty-four eigenvectors. This 4-fold or 6-fold degeneracy is disadvantageous for the ADM formulation, because these degenerate subspaces lead to linear growth of the constraint violating modes.

To see this one has to understand the structure of the 4-dimensional degenerate subspace ($A\neq 1/2$). A particularly useful tool for doing this is  a Jordan decomposition (JD) of matrix $M$. If $A=1/2$, one can perform the exact same procedure to study the six-dimensional degenerate subspace. 

The JD is a similarity transformation $M=XJX^{-1}$, where $X$ is a non-singular transformation matrix and $J$ is called the Jordan matrix, which has the eigenvalues of $M$ on its diagonal. If $J$ is diagonal then $M$ has a complete set of eigenvectors, if however $J$ is not diagonal, then the off diagonal elements (which can be placed either above or below the main diagonal) are equal to unity. These off-diagonal elements can give the information of the combination of dynamical variables, which are responsible for the fact that matrix $M$ does not have a complete set of eigenvectors. This can be seen, if the eigenvalue problem \eqref{eig_prob} is written as follows
\labeq{jord_decomp}{J (X^{-1} {\bf v})=\omega (X^{-1}{\bf v}).
}

Equation \eqref{jord_decomp} is an eigenvalue problem for matrix $J$. One can find the Jordan matrix $J$ and the non-singular transformation matrix $X$ by 
standard methods and hence one can also find which components of the transformed vector $X^{-1}{\bf v}$ correspond to the off-diagonal elements of $J$. 

By carrying out the JD in the ADM case one finds that the four zero modes, which are thought to be ``non-evolving" and  which belong to the four-dimensional degenerate subspace, are described by the following equations
\beqar
\partial_t \gamma_{12} & = & -2 K_{12}, \label{4subspace1}\\
\partial_t K_{12} & = & -\half A \partial_x D_{211}- \half A \partial_x  D_{222} \nonumber\\
				& & -\frac{1}{4}(1+2A)\partial_x D_{233}    
					+\frac{1}{4}\partial_x D_{323}, \label{4subspace2}  \\
%\eeqar
%
%\beqar
\partial_t \gamma_{13} & = & -2 K_{13}, \label{4subspace3} \\
\partial_t K_{13} & = &\ \ \frac{1}{4} \partial_x D_{223}-\half A\partial_x  D_{311} \nonumber\\
				&  & -\frac{1}{2}(1+2A)\partial_x D_{322}
					-\frac{1}{2}A\partial_x D_{333}. \label{4subspace4}
\eeqar

Plugging in the components of the zero-speed eigenvectors shows that the RHS of \eqref{4subspace2} and \eqref{4subspace4} is 0. Hence, components $\gamma_{12},K_{13},\gamma_{13},K_{13}$ should not evolve with time. This is true only if the RHS is exactly zero. However, small numerical perturbations can result in an non-zero RHS. As a result, $K_{12}$ and $K_{13}$ will begin to grow linearly in time. Non-zero values of $K_{12}$ and $K_{13}$,  in turn lead to violations of ${\cal M}_2$ and ${\cal M}_3$ respectively. This linear growth in the non-linear regime, where the perturbations are strong, is likely to obtain explosive behavior and eventually terminate simulations. This is precisely the root of the ill-posed nature of the ADM formulation.

%The results of this section are in excellent agreement with the simulations shown in \cite{PaschalidisHansen}.  

So far, even the  analysis of  flat space-time shows that one has to seek for other formulations, which at least a complete set of eigenvectors. This is exactly what was achieved by the KST formulation. \\

\begin{widetext}
\begin{center}
\begin{table} 
\begin{tabular}{ccccc}\hline\hline
\multicolumn{1}{p{2.8 cm}}{%Mode 
Number of modes\quad} & 
\multicolumn{1}{c}{Speed $v_i$ \quad} & 
\multicolumn{1}{c}{Violates ${\cal C}_{kij}$? \quad} & 
\multicolumn{1}{c}{Violates ${\cal H}$? \quad} & 
\multicolumn{1}{c}{Violates ${\cal M}_{i}$?\quad} \\ \hline\hline
%(1)-(2)
2 		  & $0$		  &	 x				         & x                             &              x                    \\  \hline
%(3)-(4)
2		 &  $\pm\sqrt{2A}$  & x	         & x                             &              x                    \\  \hline 
%(5)-(8)
4          &  $\pm	1$ & x			         & x                             &              x                    \\  \hline
%(9)-(12)
4		& $\pm 1$  &  \checkmark	&	x        &      \checkmark      \\ \hline
%(13)-(14)
2       &  $\pm 1$     &  \checkmark    &        \checkmark          &       \checkmark              \\ \hline
%(15)-(28)
14       &  $0$     &  \checkmark    &      x          &      x              \\ \hline
%(29)-(30)
2       &  $0$     &  \checkmark    &      \checkmark          &      x              \\ \hline
\end{tabular}
\caption{Classification of the KST flat space modes into constraint satisfying and constraint violating. The first column gives the number of modes. The second column gives the characteristic speed, which corresponds to the modes of the first column. The third, fourth and fifth columns have either a ``x" or a ``\checkmark" as values. A ``x" means that the constraint is not violated by the specific modes, whereas a checkmark ``\checkmark" implies that at least one of the components of the constraint is violated by the specific modes..}	
\label{table2}
\end{table}
\end{center}

\end{widetext}

%%%%%%%%%%%%%%%%%%%%%%%%%%%%%%%%%%%%%%%%%%%%%%%%%%%%%%
\subsection{Flat space stability with the KST formulation \label{flat_space_KST}}
%%%%%%%%%%%%%%%%%%%%%%%%%%%%%%%%%%%%%%%%%%%%%%%%%%%%%%

%The KST equations can be extracted from equations \eqref{Fourier_flat_space_evol1}-\eqref{Fourier_flat_space_evol3} by setting $\{\lambda=\epsilon=\xi=\phi=\theta=0\}$. 
If the KST formulation is employed, matrix $M$ of equation \eqref{eig_prob} has twelve non-zero eigenvalues and a complete set of eigenvectors, irrespectively of the value of $A$. Nevertheless, it still has sixteen zero eigenvalues which correspond to constraint violating solutions. 

A classification of the KST modes into constraint satisfying and constraint violating is given in table~\ref{table2}. Note, that the results presented in table~\ref{table2} are for the values of parameters given by equation \eqref{c2_c3_1b}.

A few remarks are worth making at this point: 
%\begin{itemize}
a) The components of the constraint  satisfying modes of the KST formulation (first 3 rows of table~\ref{table2}) are exactly the same as those of the ADM formulation independently of  whether $A=1/2$ or $A\neq 1/2$, as expected. %The constraint violating modes (rows 4-7 of table~\ref{table2}) are different.
%, because the RHS of the evolution equations of the KST formulation is not the same as that of the ADM one.

b) The KST formulation has cured the pathology of the incompleteness of eigenvectors of the ADM formulation. 
%From the results above, it appears that mode (15) and mode (20) (in the current section) become exactly zero, if $\chi=2$ and $A=1/2$, leading thus to an 
%incomplete set of twenty-eight eigenvectors. However, this is only an apparent result. If one calculates matrix $M$ and carries out the analysis all over from the 
%beginning, with $\chi=2, A=1/2$, one still finds a complete set of eigenvectors, with eigenvalues $0,\pm 1$, as expected. Another, note regards modes (13) and 
%(14). According to table \ref{table2}  these two modes violate components of all constraints. 
Note further that if $\chi=-2/5$,  the number of modes that violate the Hamiltonian constraint are  minimized. It is not known yet, whether this is true for arbitrary spacetimes. For this, a detailed study is required, which is outside the scope of this work. However, it is interesting to notice the coincidence between this property and equations \eqref{c2_c3_1d}.

c) Despite the desired property of strong hyperbolicity, the KST formulation has eighteen static modes sixteen of which violate the ${\cal C}_{kij}$ constraints. According to \cite{Alcubierre}, the existence of static modes can lead to exponential explosion caused by low-order term perturbations. 

The disadvantages of the KST formulation were demonstrated in \cite{PaschalidisHansen}, where it was shown that numerical simulations with the KST system in conjunction with finite difference methods are unexpectedly terminated in certain cases.  In particular, the KST simulations of gauge waves and expanding Gowdy spacetimes were terminated even earlier than the ADM simulations. In addition, linear growth of ${\cal C}_{kij}$ was observed in the evolution of initially small random noise. The unexpected behavior of the KST formulation was found to be caused by the growth of ``static modes", which were excited by the unavoidable small violations of the momentum constraints. 
% The suggested cure for this pathology is to extend the KST system to system 2 of section \ref{system_2}. 
%If one carries out the analysis for system 2, one will find that the ${\cal C}_{kij}$ constraint violating modes are damped according to equation \eqref{tCkijsystem2}. The analysis of system 2 will not be presented here, because the analytic expressions of the eigenvalues and eigenvectors are long, complicated and would take a lot of space to be presented. However, a careful analysis of the effect of the $\zeta$ term has been presented in \cite{Robert_Owen}, and the effect of the $\lambda$ term can be understood by the analysis of system 1, which is presented in section \ref{flat_space_system_1}. 

%\end{itemize}

%%%%%%%%%%%%%%%%%%%%%%%%%%%%%%%%%%%%
\subsection{Flat space stability with system 1 \label{flat_space_system_1}}
%%%%%%%%%%%%%%%%%%%%%%%%%%%%%%%%%%%%

%The equations of system 1 can be extracted from equations \eqref{Fourier_flat_space_evol1}-\eqref{Fourier_flat_space_evol3}
% by setting $\{\rho=\psi=\chi=\eta=0\}$. 
The flat space stability analysis with system 1, shows that system 1 is the most stable of the 3 formulations. 
For the choice of parameters that make system 1 well-posed, i.e. when both conditions \eqref{pp_cond} and \eqref{xiphi} are satisfied, and one of equations \eqref{cond1a} or \eqref{cond1b} holds true, $M$ has twenty-eight non-zero eigenvalues and a complete set of eigenvectors. 
%
%\begin{enumerate}
%\begin{itemize}
%\item[a)] $\lambda=\epsilon=-2\xi=-2\phi\neq \theta/2>0$, 
%\item[b)] $2\lambda\neq 2\epsilon=-2\xi=-2\phi=\theta/2>0.$
%\end{enumerate}
%\end{itemize}
%
%
%Condition a) implies that all damping rates are the same except for those two ($\theta/2$), which correspond to the momentum constraint violating modes (cf. section \ref{system_2}). Condition b) implies that the damping rates of the violation of the momentum constraints are equal to each other and equal to the damping rate of the hamiltonian constraint, which in turn is double the damping rate of violations of the ${\cal C}_{klij}$ constraints and different than the damping rate of ${\cal C}_{kij}$ constraints (cf. section \ref{system_2}). The value of $\zeta$ has only been found to be essential for the successful damping behavior of ${\cal C}_{kij}$, since if $\zeta<0$, exponential explosion of the ${\cal C}_{kij}$ constraints will take place after some finite time (see section \ref{model_equation}).

The results presented in this section are specifically for parameters which satisfy \eqref{cond1a} and $\zeta=0$. The analysis with parameters satisfying \eqref{cond1b} and $\zeta=0$ is qualitatively the same. The reason why $\zeta$ is set to zero, is because the results with non-zero $\zeta$ are rather complicated. The effect of the $\zeta$ term has already been discussed both in section \ref{model_equation} and in \cite{Robert_Owen} and that is the wavelength-independent damping of all modes which violate ${\cal C}_{kij}$. 
%. By simple inspection of equations \eqref{Fourier_Lin_Constraints} it can be determined whether the modes below satisfy or violate the constraints. 

In addition to the choice of parameters described above, a harmonic lapse function ($A=1/2$) has been used here for simplicity. The results remain qualitatively the same for any value of $A$. A classification of the system 1 flat space modes into constraint satisfying and constraint violating follows in table \ref{table3}.
Notice that in this case the eigenvalues have been given, instead of the characteristic speeds. 

A few remarks can be readily made: a)  The eigenvalues of all constraint violating modes  are non-zero and in particular they have negative imaginary part, which implies exponential damping. The damping timescale is inversely proportional to $\kappa^2$, which means that the short wavelength perturbations are damped extremely efficiently. 

%b) It is straightforward to check that at least in the linearized regime, the completeness of the set of eigenvectors is not important anymore. In the ADM case (see appendix \ref{flat_space_ADM}) the incompleteness of the eigenvectors results in a $n$-dimensional degenerate sub-space, to which there correspond zero eigenvalues. It is this combination of degeneracy with zero eigenvalues, which leads to linear growth. However, for system 1, even when there are less than 30 independent eigenvectors the number of non-zero eigenvalues remains the same and to any degenerate subspace there correspond exponentially damping modes.  This can be more easily realized, if one carries out a Jordan decomposition of matrix $M$ (see appendix \ref{flat_space_ADM}) for any of the cases, where there is not a complete set of eigenvectors. Then one will see that any degenerate subspace is really damped and no such thing as growth of the constraints can take place. It is concluded then, that the need for a complete set of eigenvectors is not crucial for system 1. This argument is further strengthened by the structure of equations \eqref{tCkijparabolic}-\eqref{tMiparabolic}, which remains the same even when the evolution equations of the dynamical variables do not have a complete set of eigenvectors. However, the completeness of the set of eigenvectors is desired, for then one has a basis and by virtue of that basis all modes can be decomposed and a better understanding of the behavior of the system can be achieved.

b) Another comment regards the nature of the system 1 modes. The first eight constraint-satisfying modes propagate as waves (the eigenvalues are purely real), except for two which are static. These modes are the same among all formulations, which are derived by addition of combinations of constraints to the RHS of the ADM formulation. 

All remaining modes are only damped, except for the two of the last row of table \ref{table3}. These two modes represent damped traveling waves. The origin of these modes can be understood, if one takes a closer look at the ADM modes. In table \ref{table1}, it is seen that apart from the six traveling constraint satisfying modes, there are another two traveling wave modes which violate the constraints. The parabolic terms of system 1 are second order (see equations \eqref{flat_space_evol1}-\eqref{flat_space_evol2}) and therefore they  cannot interact with the first order terms of the ADM formulation and lead to an exact cancellation of the two ADM traveling constraint-violating modes. However, because these ADM modes violate the constraints, system 1 damps them while they propagate.

c) Finally, the zero eigenvalues (first row of table~\eqref{table3}) correspond to constraint satisfying modes and thus they do not pose any threat. Nevertheless, a fixed shift $\beta^i=\mbox{constant}$ assigns  non-zero speed to these two modes. In particuarly they become equal to $\beta^i\kappa_i$, where $\kappa_i$ is the one-form which corresponds to the wavevector.

All aforementioned predictions about the damping properties of 
system 1 have been numerically confirmed in \cite{PaschalidisHansen}.

\begin{widetext}
\begin{center}
\begin{table} 
\begin{tabular}{ccccc}\hline\hline
\multicolumn{1}{p{2.8 cm}}{%Mode 
Number of modes \quad} & 
\multicolumn{1}{c}{Eigenvalue $\omega_i$ \quad} & 
\multicolumn{1}{c}{Violates ${\cal C}_{kij}$? \quad} & 
\multicolumn{1}{c}{Violates ${\cal H}$? \quad} & 
\multicolumn{1}{c}{Violates ${\cal M}_{i}$?\quad} \\ \hline\hline
%(1)-(2)
2 		  & $0$		  &	 x				         & x                             &              x                    \\  \hline
%(3)-(4)
2		 &  $\pm\kappa$  & x	         & x                             &              x                    \\  \hline 
%(5)-(8)
4          &  $\pm	\kappa$ & x			         & x                             &              x                    \\  \hline
%(9)-(10)
2		& $- i\half\epsilon\kappa^2$  &  x	 &	x        &      \checkmark      \\ \hline
%(11)-(20)
10       &  $- i\epsilon\kappa^2$     &  \checkmark    &       x          &       \checkmark              \\ \hline
%(21)-(28)
8       &  $- i\epsilon\kappa^2$     &  \checkmark    &      x          &      x              \\ \hline
%(29)-(30)
2       &  $- i\epsilon\kappa^2 \pm \kappa$     &  \checkmark    &      \checkmark          &      \checkmark              \\ \hline
\end{tabular}
\caption{Classification of the flat space modes of system 1 into constraint satisfying and constraint violating. The first column gives the number modes.
%, in the order the modes have been presented in the text. 
The second column gives the eigenvalues, which correspond to the modes of the first column. If the eigenvalue is purely real, then the corresponding mode propagates as a wave, whereas if the imaginary part of the eigenvalue is negative, the corresponding mode is getting damped. The third, fourth and fifth columns have either a ``x" or a ``\checkmark" as values. A ``x" means that the constraint is not violated by the specific modes, whereas a checkmark ``\checkmark" implies that at least one of the components of the constraint is violated by the specific modes.}	
\label{table3}
\end{table}
\end{center}

\end{widetext}

\section{On the choice of parameters \label{parameters}}

This section discusses some ideas, which are related to the choice of parameters of system 1 and 2. It should be noted that this is a complicated discussion and only preliminary arguments will be made which are by no means complete. An analysis of this delicate subject has also been given in \cite{PaschalidisHansen}. 
%

%Both system 1 and system 2 are parabolized extentions of hyperbolic formulations and for this reason the discussion below regards both systems. 

Because the equations have been parabolized, the values of the free parameters depend mainly on two factors: a) The effectiveness of the damping rate and b) the numerical scheme employed to integrate the evolution equations. 

As already mentioned in sections \ref{model_equation} and \ref{paraADM_con_evol}, the damping rate is of the form $\exp(-\alpha\lambda\kappa^2t)$. Therefore, what one would desire is for the minimum damping rate to be efficient enough.
The minimum damping rate is $\exp(-\alpha_{min}\lambda\kappa_{min}^2t)$, where $\alpha_{min}$ is the minimum value of the lapse function and $\kappa_{min}=2\pi/L$ with $L$ the length of the computational domain. Therefore, if  $\alpha_{min}$ is very small then the damping rate is very small, unless the system 1 parameters are assigned large values.

In principle, setting the damping parameters to very large values results in a very efficient drive towards the surface of constraints. Nevertheless, there are numerical issues involved with parabolic equations which have to be addressed. Before the validity of the following discussion is taken for granted a Von Neuman stability analysis \cite{Morton} of the evolution equations is required. However, general statements can still be made even if such an analysis is not available. 

The numerical stability of an explicit numerical scheme for parabolic equations requires that the mesh size and damping parameter satisfy a Courant condition of the form $\lambda \Delta t/(\Delta x)^2\leq c$, where $\Delta t$ is the time step, $\Delta x$ is the size of the mesh and c is a constant. For example, $c=1/2$ for a one dimensional linear diffusion equation with a four point explicit scheme with forward time difference \cite{Morton}. 

The above Courant condition implies that the time step has to be really small, if $\lambda$ is large. It is this tradeoff between time marching versus strong damping, which is the limiting factor on the values of the damping parameters with explicit numerical schemes. 

Similar constraints on the maximum time step exist when explicit schemes are chosen for the ADM evolution equations. In this case, it is the speed of propagation of the information, which sets the limit on the time step. For most applications, numerical stability with the ADM formulation requires a Courant condition of the form $\mu=\Delta t/\Delta x\leq \tilde c$, where $\mu$ is the usual Courant number. The same is true for other hyperbolic systems. For example, the numerical integration of a one-dimensional wave equation with characteristic speed unity and a standard explicit scheme \cite{Morton}  requires $\mu=\Delta t/\Delta x\leq 1$. 

Systems 1 and 2 contain both parabolic and hyperbolic terms, and hence for numerical stability both the parabolic and the hyperbolic condition have to be satisfied \cite{Morton}. Combining the two conditions, one finds that for system 1 in conjunction with explicit schemes the overall numerical stability is of the form $\lambda\mu\leq \bar c\Delta x$. Thus, $\lambda$ should have a value, such that $\mu$ and $\Delta x$ should be large enough in order to finish computations within reasonable time-scales, while maintaining satisfactory damping properties. This has been shown to be possible in \cite{PaschalidisHansen}.

Note, however, that the above limitation on the time-step is not present when the numerical integration is carried out with implicit schemes. For example, a Crank Nickolson numerical scheme for the one-dimensional diffusion equation \cite{Morton} is always stable. An implicit scheme would allow one to set the damping parameters to arbitrarily large values without having to worry about limitations on the grid structure, but at the cost of cumbersome programming.  

%The equations with the parabolic terms have turned so complicated that perhaps only via numerical experiments one may eventually be in position to make 
%quantitative statements. 
Taking all these facts into account, it is concluded that until more detailed numerical simulations become available the safe choice for the parameters of systems 1 and 2 are those which satisfy the conditions for well-posedness presented in sections~\ref{wp_system_1} and ~\ref{sys2_wp} respectively.  
%
%\labeq{parameter_choice1}{
%\lambda>0,\epsilon>0,\zeta>0,\theta>0,\phi<0,\xi<0
%}
%
%The aforementioned choices satisfy the necessary conditions and all the conditions for system 2 to be well-posed.

%%%%%%%%%%%%%%%%%%%%%%%%%%%%%%%%%
\section{Conclusions \label{discussion}}
%%%%%%%%%%%%%%%%%%%%%%%%%%%%%%%%%

Two new parameterized 3+1 formulations have been presented. These two formulations are referred to as
system 1 and system 2. 

The well-posed system 1 is derived by adding combinations of the constraint equations and their derivatives 
to the RHS of the first-order ADM formulation. As a result the evolution equations of the dynamical variables have become second order in spatial derivatives and it has been shown that system 1 has obtained mixed hyperbolic - second-order parabolic nature. 

The evolution of the constraints with system 1 is governed by 2nd-order parabolic PDEs irrespectively of the gauge condition employed. This in turn means that the constraints are damped and smoothed in time, and hence the surface of constraints is a local attractor.

The well-posed system 2, is a parameterized parabolization of what is usually called  
the KST formulation. The original KST gauge condition was generalized and a certain combination of the terms of system 1 was added to the RHS of the KST formulation. System 2 is a manifestly mixed hyperbolic - second-order parabolic  system of equations.  

The effect of the terms introduced in system 2 is the efficient damping and smoothing of perturbations which violate a subset of the constraints of the formulation, while the remaining constraints evolve according to strongly hyperbolic equations. 
%An important advantage of system 2 is that it is relatively easy to integrate with the KST system of redefined variables (see section \ref{system_2}) as it was 
%presented in \cite{KST}. This system of redefined variables has demonstrated better stability properties than the untransformed KST formulation. Therefore, it 
%may be more advantageous to adapt system 2 to the transformed KST formulation instead. 

%Note also, that  the parabolic terms of system 1 might as well be added to the RHS of the KST formulation, but then a detailed analysis is required in order to understand how the parabolic terms interact with the hyperbolic terms of the KST formulation. This will be the subject of future work.

As a first test for system 1, a study of the stability of flat space against small amplitude perturbations was carried out and contrasted to the stability of the ADM and the KST formulations. The results indicate that in this case system 1 is more stable than both the ADM and the KST formulations. In particular, it was shown that system 1 damps exponentially and smooths all finite wavelength constraint-violating modes. Numerical support for this prediction  has been given in \cite{PaschalidisHansen}.

These two new formulations provide a new basis for obtaining well-posed formulations of GR which possess constraint damping and in turn a new basis
for constructing numerical schemes for long-term and stable numerical simulations of the Einstein equations.

\acknowledgements

I thank Jakob Hansen for reading the manuscript and for his constructive comments and Alexei Khokhlov, for extensive discussions and advice. Finally, it is a pleasure to thank Carlo Graziani for attracting my attention to the possibility of using higher order systems of partial differential equations in numerical relativity.

%%%%%%%%%%%%%%%%%%%%%%%%%%%%%%%%%%%%%%%%%%%%%%%%%%%%%%%%%%%%%%%%%%%%%%%%%%%%%%%%%%%%
%%%%%%%%%%%%%%%%%%%%%%%%%%%%%%%%%%%%%%%%%%%%%%%%%%%%%%%%%%%%%%%%%%%%%%%%%%%%%%%%%%%%
%%%%%%%%%%%%%%%%%%%%%%%%%%%%%%%%%%%%%%%%%%%%%%%%%%%%%%%%%%%%%%%%%%%%%%%%%%%%%%%%%%%%

\appendix

%%%%%%%%%%%%%%%%%%%%%%%%%%%%%%%%%%%%%%%%%%%%%%%%
\section{Evolution of the Constraint Equations \label{appA}}
%%%%%%%%%%%%%%%%%%%%%%%%%%%%%%%%%%%%%%%%%%%%%%%%

In this appendix the approach of \cite{YonedaShinkai} is followed and  the evolution equations of the constraints of the first order ADM formulation (not of the standard ADM formulation as was done in \cite{YonedaShinkai}) are given in terms of the time derivatives of the dynamical variables instead of in terms of the spatial derivatives of the constraints and the constraints themselves (cf. equations \eqref{tCkij}-\eqref{tM}). This is very useful because the addition of constraints to the RHS of the ADM formulation has a direct impact on the evolution equations of the constraints, which becomes straightforward to calculate if the evolution equations of the constraints are expressed in terms of the time derivatives of the dynamical variables.

It is straightforward to check that by taking the time-derivative $\dt$ of the constraint equations of the first order ADM formulation one obtains

\labeq{tCkij_dyn}{
\dt {\cal C}_{kij}=\partial_k(\dt \gamma_{ij})-(\dt D_{kij}),}
\labeq{tCklij_dyn}{\begin{split}
\dt  {\cal C}_{klij}= \partial_k(\dt D_{lij})-\partial_l(\dt D_{kij}) 
\end{split},}
\begin{widetext}
\labeq{tH_dyn}{\begin{split} 
\dt  {\cal H}= & ^{(1)}H^{skij}\partial_s(\dt D_{kij})+\ ^{(2)}H^{kij}(\dt D_{kij})+\ ^{(3)}H^{ij}(\dt K_{ij})+\ ^{(4)}H^{ij}(\dt \gamma_{ij}),
\end{split}
 }
where 
\labeq{H1_4}{\begin{split}
^{(1)}H^{skij}= &\ \gamma^{kj}\gamma^{si}-\gamma^{ij}\gamma^{ks}, \\
^{(2)}H^{kij}= &\ \frac{3}{2}\gamma^{ij}\gamma^{mn}\gamma^{ks}D_{mns}-\half\gamma^{ij}\gamma^{mn}\gamma^{ks}D_{smn}-
2\gamma^{ki}\gamma^{mn}\gamma^{js}D_{mns} \\
			&\ + \gamma^{ki}\gamma^{sj}\gamma^{mn}D_{smn}+\frac{3}{2}\gamma^{mi}\gamma^{nj}\gamma^{lk}D_{lmn}-
	\gamma^{li}\gamma^{sj}\gamma^{km}D_{sml}, \\
^{(3)}H^{ij}= & \ 2K\gamma^{ij}-2K^{ij}, \\
^{(4)}H^{ij}= & \ 2 K^{in}K_{n}{}^j-2R^{ij}-2KK^{ij}+\gamma^{mn}\Gamma^i{}_{mn}\gamma^{ks}\Gamma^j{}_{ks} 
			  -\gamma^{ia}\gamma^{jb}\Gamma^{s}{}_{ak}\Gamma^k{}_{sb} \\
  			& \ -\gamma^{ia}\gamma^{jb}\gamma^{sl}{\cal C}_{asbl} 
			 -\gamma^{si}\gamma^{mj}\gamma^{ln}\gamma^{ab}D_{amn}D_{bsl}
			+\gamma^{si}\gamma^{nj}\gamma^{mk}\gamma^{ab}D_{ank}D_{smb},
\end{split}
}
\labeq{tM_dyn}{\begin{split}
\dt {\cal M}_i=&\ ^{(1)}M_i{}^{kab}\partial_k(\dt K_{ab})+\ ^{(2)}M_i{}^{ab}(\dt K_{ab})+\ ^{(3)}M_i{}^{kab}(\dt D_{kab})+
				\ ^{(4)}M_i{}^{ab}(\dt \gamma_{ab}),
\end{split}
 }
\end{widetext}
where
\labeq{M1_4}{\begin{split}
^{(1)}M_i{}^{kab} = & \  \gamma^{ka}\delta_i{}^b-\gamma^{ab}\delta_i{}^k,       \\
^{(2)}M_i{}^{ab}= &\ \gamma^{an}\Gamma^b{}_{in}-\gamma^{mn}\Gamma^a{}_{mn}\delta_i{}^b, \\
^{(3)}M_i{}^{kab}= & \ \half\gamma^{ab}K_i{}^k-\gamma^{ka}K_i{}^b+\half K^{ab}\delta_i{}^k,  \\
^{(4)}M_i{}^{ab} = & \ \nabla_i K^{ab}-\nabla^a K^b{}_i+\gamma^{mn}\Gamma^{a}{}_{mn} K_i{}^b-K^{an}\Gamma^b{}_{in}.
\end{split}
}
\\
Note that quantities $^{(1)}M_i{}^{kab}$, $^{(2)}M_i{}^{ab}$, $^{(3)}M_i{}^{ij}$, $^{(4)}M_i{}^{ab}$, $^{(1)}H^{skij}$, $^{(2)}H^{kij}$, $^{(3)}H^{ij}$ and $^{(4)}H^{ij}$ here, should not be confused with the similar objects $H^{mns}$, $^{(1)}M^{msn}$,  $^{(2)}{M}_i{}^{msn\ell}$,  $^{(3)}M^{mn}$ and $^{(4)}M_i{}^{msn}$ in equations \eqref{tCkij}-\eqref{tM}.

A final note concerns the $^{(4)}H^{ij}$ object. By a closer look, one realizes that the constraint ${\cal C}_{klij}$ is part of a term in the definition of $^{(4)}H^{ij}$. This means that addition of constraints to the RHS of the first-order ADM evolution equations, will result in evolution equations for the constraints, which will have quadratic dependence on the constraints. This resulted, because in the derivation of \eqref{tCkij_dyn}-\eqref{M1_4}  the Ricci tensor has not been calculated through \eqref{Riccimodified}. If \eqref{Riccimodified} is enforced, then the ${\cal C}_{klij}$ term vanishes from the expression of $^{(4)}H^{ij}$, but the expressions of the equations above have to change accordingly. 

%%%%%%%%%%%%%%%%%%%%%%%%%%%%%%%%%%%%%%%%%%%%%%%%%%%%%%%%%%%%%%%%%%%%%%%%%%%%%%%%%%%%%%%%%%
\section{Well-posedness of systems which satisfy the Petrovskii condition 
\label{app_B}}
%%%%%%%%%%%%%%%%%%%%%%%%%%%%%%%%%%%%%%%%%%%%%%%%%%%%%%%%%%%%%%%%%%%%%%%%%%%%%%%%%%%%%%%%%%

This appendix provides conditions for well-posedness of the initial value problem with quasi-linear systems of PDEs which satisfy the Petrovskii condition. 

Let 
\labeq{PDE}{\partial_t u=\tilde P(u,\partial_i) u, \ \ \ i=1,\ldots s}
be a set of quasi-linear PDEs, where $u$ is the column vector of $n$ dynamical variables, $\tilde P$ is a quasi-linear differential operator, $\partial_i$ denotes a spatial derivative and $s$ is the number of spatial dimensions. For well-posedness $\tilde P$ can be considered to coincide with its linearized and localized principal part  \cite{PDE_textbook}, which is denoted by $P$. 

Before the proposition of this appendix and its proof are presented it is important to introduce a definition and a theorem from \cite{PDE_textbook}. 

\paragraph{Definition 3:} An operator, $P$, is called semi-bounded if there exists a Hermitian operator such that $(u,Pu)_H+(Pu,u)_H\leq 2c(u,u)_H$, for all sufficiently smooth functions $u(x)$ and $c \in \mathbf{R}$ is a constant. 

The quantity $(u,w)_H$ above is called an energy inner product and its exact definition can be found in \cite{PDE_textbook}.

\paragraph{Theorem 3:} If the operator $P$ is semi-bounded the Cauchy problem for \eqref{PDE} is well-posed. 

Now the following proposition can be proven. 
\paragraph{Proposition 1:} \textit{If the Fourier transform, $P(i\kappa)$, of $P$ a) satisfies the Petrovskii condition, b) has a complete set of eigenvectors, and c) the eigenvectors can be chosen  as analytic functions for all $\kappa\neq 0$, then equation \eqref{PDE} admits a well-posed Cauchy problem.}

\textit{Proof:} Proving well-posedness in the case $\kappa=0$ is always trivial because then $P(i\kappa)=0$ \cite{PDE_textbook}. For $\kappa\neq 0$, if  $P(i\kappa)$ satisfies the Petrovskii condition then the eigenvalues $\omega_i$ of $P(i\kappa)$ satisfy \labeq{Petrovskii}{Re(\omega_i)\leq c,} where $c \in \mathbf{R}$ and is independent of $\kappa$. 

If $P(i\kappa)$ has a complete set of eigenvectors, then the matrix $S(\kappa)P(i\kappa)S(\kappa)^{-1}$ is diagonal with the eigenvalues of $P(i\kappa)$ on the diagonal, provided that $S(\kappa)^{-1}$ contains these eigenvectors as columns.  

Now, the hermitian matrix $H(\kappa)=S^*S$, where $S^*$ is the complex conjugate transpose of $S$, satisfies 
%\begin{widetext}
\labeq{inequality}{\begin{split}
HP+P^*H=& \ S^*SP+P^*S^*S, \\
										= & \ S^*\bigg(SPS^{-1}+(S^*)^{-1}P^*S^*\bigg)S, \\
										= & \ S^*\left(
\begin{array}{ccc}
\omega_1+\bar\omega_1 & 0 & \ldots  \\
0  & \ddots &  \\
\vdots & & \omega_n+\bar\omega_n
\end{array}\right)S, \\
 \leq &\  2cH,
\end{split}
}
%\end{widetext}
where $\bar\omega_i$ denotes the complex conjugate of $\omega_i$.
The definition of inequalities between Hermitian matrices can be found in \cite{PDE_textbook}. 

%Let $\mathbf{H}$ be the operator that maps 
%\labeq{}{(\mathbf{H}u)(x)=\frac{1}{(2\pi)^{s/2}}\int_{\mathbf{R}^s} e^{i\kappa_i x^i} H(\kappa) \tilde u(\kappa) d\kappa.} 
The energy inner product $(u,w)_H$ exists because of assumption c) \cite{PDE_textbook} of proposition 1. 
%Define also the $L_2$ inner product 
%\labeq{}{(u,\mathbf{H}u)=\frac{1}{(2\pi)^{s/2}}\int_{\mathbf{R}^s} e^{i\kappa_i x^i} \tilde u(\kappa)^* H(\kappa) \tilde u(\kappa) d\kappa.} 
Then, the following energy estimate follows
%\vspace{0.9cm}
%\begin{widetext}
\labeq{inequality_quasi}{\begin{split}
 (u,Pu)_H+(Pu,u)_H = & \\ 
			&\hspace{-1cm} = (P(i\kappa)\tilde u,H(\kappa)\tilde u)+(\tilde u,H(\kappa)P(i\kappa)\tilde u), \\ 
			&\hspace{-1cm} = \bigg(\tilde u, (H(\kappa) P(i\kappa)+P(i\kappa)^* H(\kappa))\tilde u\bigg), \\
			&\hspace{-1cm} \leq  \ 2 c (u,u)_H,
\end{split}}
%			&\hspace{-1cm} 
%				= & (P(i\kappa)\tilde u,H(\kappa)\tilde u)+(\tilde u,H(\kappa)P(i\kappa)\tilde u), \\ 
%			&\hspace{-1cm}
%			 = & \bigg(\tilde u, (H(\kappa) P(i\kappa)+P(i\kappa)^* H(\kappa))\tilde u\bigg), \\
%			&\hspace{-1cm}
%			 \leq  \ 2 c (u,u)_H,
%\end{widetext}
%
where $\tilde u$ is the Fourier transform of $u$. Parseval's relation, $(u,w)_H=(\tilde u, \tilde w)_H$, has been used in lines 1 and 3 and inequality \eqref{inequality} has been used in going from line 2 to line 3 above. 

Inequality \eqref{inequality_quasi} proves that $P$ is a semi-bounded operator.  Then according to theorem 3,  \eqref{PDE}
admits a well-posed initial value problem. This completes the proof of proposition 1.

% Note that the proven proposition guarantees only local existence of solutions as it is the case with most non-linear systems of PDEs.
% \cite{PDE_textbook}. 

%%%%%%%%%%%%%%%%%%%%%%%%%%%%%%%%%%%%%%%%%%%%%%%%%%%%%%%%%%%%%%%%%%%%%%%%%%%%%%%%%%%%%%%%%%
\section{Fourier Transform of linearized evolution and constraint equations of ADM, KST and system 1
\label{app_C}}
%%%%%%%%%%%%%%%%%%%%%%%%%%%%%%%%%%%%%%%%%%%%%%%%%%%%%%%%%%%%%%%%%%%%%%%%%%%%%%%%%%%%%%%%%%

This appendix presents the Fourier transform of equations \eqref{flat_space_evol1}-\eqref{Lin_Constraints}.

The Fourier transformed evolution equations are
\labeq{Fourier_flat_space_evol1}{\begin{split}
\omega \tilde\gamma_{ij}=& -2i\delta K_{ij}-i\lambda\kappa^2\tilde\gamma_{ij}+\lambda\kappa D_{1ij}, \\
\end{split}}
\labeq{Fourier_flat_space_evol2}{\begin{split}
\omega \tilde K_{ij}= 
%ADM
				& -\frac{1}{4}\kappa[\tilde D_{ji1}+\delta_{1j} \tilde D_{kik}+\tilde D_{ij1}+ \\
					&+ \delta_{1i}\tilde D_{kjk}-(1+2A)\delta_{1i}\tilde D_{jkk}- \\
					& -(1+2A)\delta_{1j}\tilde D_{ikk}
					-2 \tilde D_{1ij}] + \\
% KST
					& -\rho\delta_{ij}\kappa(D_{mm1}-D_{1mm}) \\
					& -\psi\kappa( D_{(ij)1}-\delta_{1(i} D_{a j)a}) \\
%
%PARABOLIC ADM
					&+ i\phi\kappa^2\delta_{ij}(\tilde K_{22}+\tilde K_{33})- \\
					&- i\half\theta\kappa^2(\delta_{1j}\tilde K_{1i}+\delta_{1i}\tilde K_{1j})+ \\
					&  +i\theta\kappa^2\delta_{1i}\delta_{1j}(\tilde K_{11} +\tilde K_{22}+
					   \tilde K_{33}), \\
\end{split}
}
\labeq{Fourier_flat_space_evol3}{\begin{split}
\omega \tilde D_{1ij} = & 2 \kappa\tilde K_{ij}
% KST        
				 -\kappa\eta \big(\delta_{1(i}K_{j)1}-\delta_{1(i}\delta_{j)1} K_{mm}\big) \\
				& +\kappa\chi\delta_{ij}(K_{22}+K_{33}) 
%
%PARABOLIZED ADM
 -i\xi\kappa^2\delta_{ij}(\tilde D_{m1m}-\tilde D_{1mm}), \\
%%%%%
\omega\tilde D_{2ij} = 
%KST
				& -\kappa\eta \big(\delta_{2(i}K_{j)1}-\delta_{2(i}\delta_{j)1} K_{mm}\big) \\				& - \chi\delta_{ij} K_{12}
% PARABOLIZED ADM
%
 -i\epsilon\kappa^2\tilde D_{2ij}, \\
\omega\tilde D_{3ij} = 
%KST
				& -\kappa\eta \big(\delta_{3(i}K_{j)1}-\delta_{3(i}\delta_{j)1} K_{mm}\big) \\				& - \chi\delta_{ij} K_{13}
% PARABOLIZED ADM
%
 -i\epsilon\kappa^2\tilde D_{3ij}. \\
\end{split}
}
The Fourier transformed linearized constraint equations are 
\labeq{Fourier_Lin_Constraints}{
\begin{split}
\tilde {\cal C}_{1ij} \equiv &\ i\kappa\tilde \gamma_{ij}-\tilde D_{1ij} =0,\\
\tilde {\cal C}_{2ij}\equiv &\   \tilde D_{2ij} =0,\\
\tilde {\cal C}_{3ij}\equiv &\    \tilde D_{3ij} =0,\\
\tilde {\cal H}\equiv &\   \tilde D_{212}+\tilde D_{313}- \tilde D_{122}-\tilde D_{133}=0, \\
\tilde {\cal M}_1 \equiv &\   \tilde K_{22}+\tilde K_{33} =0, \\
\tilde {\cal M}_2 \equiv &\   \tilde K_{12} =0,\\
\tilde {\cal M}_3 \equiv &\   \tilde K_{13} =0.
\end{split}
}

Note that in \cite{Alcubierre} it was shown that only three of the four constraints (Hamiltonian and Momentum) are independent, whereas in \eqref{Fourier_Lin_Constraints}  all four are independent. The apparent discrepancy arises, because the first-order ADM evolution equations are not reduced to equations for the three-metric only. This is exactly what was done in \cite{Alcubierre}, and hence the analysis presented in \cite{Alcubierre} was for a different set of equations than those of the first-order ADM formulation.  

The reason why this approach is not followed in this work is that the totality of the modes present in the first-order ADM formulation is larger than that of the second-order ADM formulation. The first-order ADM formulation has opened up an eighteen-dimensional subspace, where more instabilities than the standard ADM formulation may arise. The totality of solutions of the first order ADM is greater than that of the standard second-order ADM and the two coincide, if and only if the ${\cal C}_{kij}$ constraints are satisfied \cite{Courant}. The proposition in this work is to simulate the Einstein equations using variants of the first-order ADM formulation. Therefore, a study of the totality of the modes present in the first-order ADM formulation and its variants is required and the evolution equations will not be reduced to equations for the three metric only.  %\vspace{2cm}

%\newpage

%%%%%%%%%%%%%%%%%%%%%%%%%%%%%%%%%%%%%%%%%%%%%%%%%%%%%%%%%%%%%%%%%%%%%%%%%%%%%%%%%%%%%%%%%%%%%%%%%%%%%%%%%%%%
%%%%%%%%%%%%%%%%%%%%%%%%%%%%%%%%%%%%%%%%%%%%%%%%%%%%%%%%%%%%%%%%%%%%%%%%%%%%%%%%%%%%%%%%%%%%%%%%%%%%%%%%%%%%

\end{document}